\pdfoutput=1 
\documentclass{jfm}
\usepackage{graphicx}
\usepackage{epstopdf, epsfig}
\usepackage{hyperref}

\shorttitle{Non-uniform EOF of power-law fluid in a Hele-Shaw cell}
\shortauthor{E. Boyko, M. Bercovici and A. D. Gat}

\title{Flow of Power-Law Liquids in a Hele-Shaw Cell Driven by Non-Uniform Electroosmotic Slip in the Case of Strong Depletion}

\author{E. Boyko,
M. Bercovici \corresp{\email{mberco@technion.ac.il}}
 \and A. D. Gat \corresp{\email{amirgat@technion.ac.il}}}
\affiliation{ Faculty of Mechanical Engineering, Technion - Israel Institute of Technology, Haifa 3200003, Israel}
\begin{document}
\maketitle
\begin{abstract}
We analyze flow of non-Newtonian fluids in a Hele-Shaw cell, subjected to spatially non-uniform electroosmotic slip. Motivated by their potential use for increasing the characteristic pressure fields, we specifically focus on power-law fluids with wall depletion properties. We derive a p-Poisson equation governing the pressure field, as well as a set of linearized equations representing its asymptotic approximation for weakly non-Newtonian behavior. To investigate the effect of non-Newtonian properties on the resulting fluidic pressure and velocity, we consider several configurations in one- and two-dimensions, and calculate both exact and approximate solutions. We show that the asymptotic approximation is in good agreement with exact solutions even for fluids with significant non-Newtonian behavior, allowing its use in the analysis and design of microfluidic systems involving electro-kinetic transport of such fluids. 
\end{abstract}
\section{Introduction}
Electroosmotic flow (EOF) is the bulk fluid motion due to an electric body force acting on the net charge in the diffuse part of an electrical double layer (EDL) near a solid surface. EOF actuation is not limited to Newtonian fluids and can be found in various microfluidic applications involving the use of complex fluids, such as ones containing proteins, colloidal suspensions, nucleic acids, or other polymeric solutions characterized by non-Newtonian behavior \citep{zhao2013electrokinetics}. In addition, recent studies have shown that the pressure field induced due to EOF is increased by two to three orders of magnitude when a non-Newtonian polymer solution is used instead of a Newtonian fluid \citep{berli2010output}. Such increase in the pressure field induced by EOF may be desirable for various applications such as more effective EOF-based pumps \citep{berli2010output} and deformation of elastic structures \citep{rubin2016}.

Non-Newtonian flow often include a depletion layer, which is a narrow region in the vicinity of the wall in which key components in the liquid (typically polymer chains) are rejected either due to the local isotropy of a Brownian motion or the electrostatic force. Within the depletion layer, the reduction in concentration of polymer segments significantly reduces the fluid viscosity, as compared to the bulk \citep{barnes1995review}.

Previous studies on EOF with depletion layer include \citet{olivares2009eof} who accounted for the effect of the depletion layer \citep{berli2008electrokinetic} on uniform EOF. The authors experimentally showed that in the case of strong depletion (where the fluid can be considered Newtonian in the vicinity of the wall), the electroosmotic velocity remains linear with applied electric field and may continue to be described by the Helmholtz-Smoluchowski slip velocity \citep{hunter2001foundations}. \citet{berli2010output} used a power-law model to examine EOF of polymer solutions in the presence of wall depletion, deriving an analytical expression for the maximal pressure difference in a capillary, and demonstrating good agreement with the experimental 
data of \citet{ paul2008electrokinetic} for polyacrylic acid.  

In many naturally occurring configurations, as well as in engineered applications, surfaces exhibit a non-uniform surface charge or zeta
potential distribution. The effects of such non-uniform distribution on the EOF of Newtonian fluids have been studied extensively in various
geometries. A number of previous studies have demonstrated analytically \citep{anderson1985electroosmosis,ajdari1995electro,ajdari1996generation,ajdari2001transverse,stroock2000patterning,qian2002chaotic,erickson2002influence,erickson2003three,zhang2006electro} and experimentally \citep{stroock2000patterning} that a periodic
zeta potential leads to multi-directional or circular flow patterns, depending on whether the applied field is parallel or perpendicular to gradient of zeta potential. Recently, we have studied EOF in a Hele-Shaw configuration with non-uniform zeta potential distribution and demonstrated that leveraging this non-uniformity enables to generate
desired flow patterns in confined regions, without physical walls \citep{boyko2015flow}.

To date, only few studies examined EOF of non-Newtonian fluids with
non-homogeneous zeta potential distributions. Recently, \citet{ng2014electroosmotic} and \citet{qi2015electroosmotic1} studied analytically and numerically EOF of power-law fluids between undulating plates with a periodic surface charge distribution. \citet{ng2014electroosmotic} showed that for a polymeric solution exhibiting strong depletion near flat surfaces, the net flux resulting from a sinusoidal zeta potential distribution is independent of its amplitude. In the absence of a depletion layer, \citet{qi2015electroosmotic1} demonstrated that increasing the amplitude of zeta potential modulation results in an increase in flux for a shear-thinning fluid and a decrease for a shear-thickening fluid. \citet{ghosh2015electroosmosis} focused on EOF of viscoelastic fluids, and also examined the case of a periodic zeta
potential distribution. Assuming a thin EDL and small Deborah number, \citet{ghosh2015electroosmosis} derived a modified Helmholtz-Smoluchowski
slip boundary for quasi-linear upper-convected Maxwell (UCM) fluids \citep{irgens2014rheology} in terms of an asymptotic series of Deborah number. The authors also showed that the net flux in the channel decreases for UCM fluids, as compared to the Newtonian ones. 

The aim of this work is to analyze the flow and pressure field of
a non-Newtonian power-law fluid within a Hele-Shaw cell, driven by
non-uniform EOF in the case of strong depletion. In $\mathsection$2 we present the problem formulation and
detail the key assumptions used in the derivation of the model. In
$\mathsection$3 we derive a non-linear p-Poisson equation governing
the pressure field, and present an asymptotic approximation for weakly
non-Newtonian power-law fluids. In $\mathsection$4 we consider a
one dimensional configuration and quantitatively estimate the asymptotic accuracy relative to the exact solution. In $\mathsection$5 we utilize
the asymptotic approximation and examine several two dimensional configurations
and characterize the effect of shear thinning and thickening on the velocity
and pressure fields. We conclude with a discussion of the results in $\mathsection$6.

\section{Problem formulation}

We study the steady non-uniform EOF of a non-Newtonian fluid within
the narrow gap between two parallel plates. Figure 1 presents a schematic
illustration of the configuration and coordinate system. We hereafter
denote dimensional variables by tildes, normalized variables without
tildes and characteristic values by an asterisk superscript. 

We employ a Cartesian coordinate system $(\tilde{x},\tilde{y},\tilde{z})$
whose $\tilde{x}$ and $\tilde{y}$ axes lie in the mid-plane of the
channel and $\tilde{z}$ is perpendicular thereto. The lower and upper
plates have arbitrary zeta potential distribution defined as $\tilde{\zeta}(\tilde{x},\tilde{y})$,
which vary over a characteristic length scale $\tilde{l}$ in the
$\tilde{x}-\tilde{y}$ plane, and the gap between the plates is $2\tilde{h}$.
Hereafter, we adopt the $\Vert$ and $\bot$ subscripts to denote parallel
and perpendicular directions to the $\tilde{x}-\tilde{y}$ plane,
respectively. The fluid constant density is $\tilde{\rho}$, fluid
velocity is $\mathbf{\tilde{\mathrm{\mathit{\boldsymbol{u}}}}}=\left(\boldsymbol{\tilde{u}}_{\Vert},\boldsymbol{\tilde{u}}_{\bot}\right)=\left(\tilde{u},\tilde{v},\tilde{w}\right)$
and fluid pressure is $\tilde{p}$. The uniform electric field, applied
parallel to the plates, is $\tilde{\mathit{\mathrm{\boldsymbol{\mathit{E}}}}}_{||}$.

\begin{figure}
 \centerline{\includegraphics[scale=1.0]{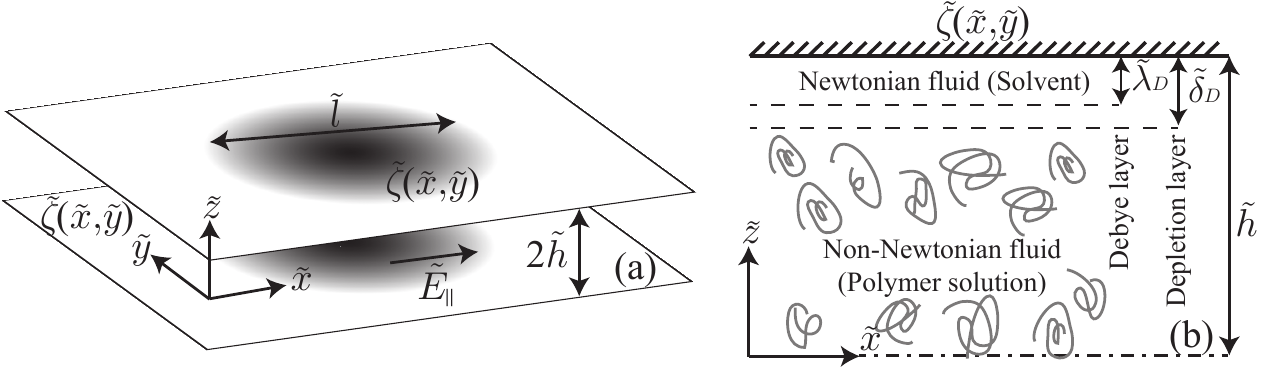}}
\caption{Schematic illustration of the examined configuration, showing the
coordinate system and relevant physical parameters. (a) Two parallel
plates contain a non-Newtonian fluid subjected to a uniform electric
field $\tilde{\mathit{\mathrm{\boldsymbol{\mathit{E}}}}}_{||}$. The
plates are separated by a small gap $2\tilde{h}$̃ and are functionalized
with arbitrary zeta potential distribution, $\tilde{\zeta}(\tilde{x},\tilde{y}).$
(b) Cross section view showing the case of strong depletion considered
in this study (dashed-dot line presents the symmetry line of the configuration).
The bulk contains a non-Newtonian fluid (polymer solution), while
the depletion layer of thickness $\tilde{\delta}_{D}$ contains a
Newtonian fluid (solvent). We assume the Debye length $\tilde{\lambda}_{D}$
is smaller than $\tilde{\delta}_{D}$.}
\end{figure}

Modelling EOF of non-Newtonian fluids may be achieved by utilizing
several constitutive models, which can be primarily divided into inelastic
power-law models \citep{das2006analytical,zhao2008analysis,berli2008electrokinetic,olivares2009eof,tang2009electroosmotic,berli2010output,zhao2010nonlinear,vasu2010electroosmotic,zhao2011exact,babaie2011combined,vakili2012electrokinetically,zhao2013electroosmotic} and viscoelastic
models, the most common being are the PTT (Phan-Thien-Tanner) model
and FENE-P (finitely extensible non-linear elastic Peterlin’s approximation)
model  \citep{afonso2009analytical,dhinakaran2010steady,afonso2011electro,sousa2011effect,afonso2013analytical}. Of all models, the commonly
used power-law model shows both experimental relevance for the EOF
problem \citep{olivares2009eof,berli2010output} and at the same time is sufficiently
simple to allow analytical treatment of non-uniform EOF. Therefore,
throughout this work we consider the following constitutive model
for the viscosity $\tilde{\eta}$ \citep{bird1987dynamics}
\begin{equation}
\tilde{\eta}(\dot{\gamma})=\tilde{\mu}_{eff}\dot{\gamma}^{n-1},\label{eq:Power Law Fluid}
\end{equation}
where $n$ is dimensionless power-law index, $\tilde{\mu}_{eff}$
is the effective constant viscosity with units $\mathrm{Pa\,s^{n}}$
and $\dot{\gamma}$ is the shear rate defined as 
and $\dot{\gamma}$ is the shear rate defined as $\dot{\gamma}=\sqrt{2\boldsymbol{\tilde{D}}:\boldsymbol{\tilde{D}}}$, where
$\boldsymbol{\tilde{D}}$  is the rate of deformation tensor given
by $\boldsymbol{\tilde{D}}=(\tilde{\boldsymbol{\nabla}}\tilde{\boldsymbol{u}}+\boldsymbol{\tilde{\nabla}\tilde{u}^{\dagger}})/2.$
The power-law index lies in range $0<n<1$ for shear-thinning fluids,
and $n>1$ for shear-thickening fluids. The case $n=1$ represents
the Newtonian fluid, in which $\tilde{\mu}_{eff}=\tilde{\eta}_{N}$,
where $\eta_{N}$ is the viscosity of the Newtonian fluid. We note
that while most polymer solutions are shear-thinning \citep{bird1987dynamics}, our analysis holds for both shear-thinning and shear thickening
fluids.

We here consider a configuration having a thin EDL, $\tilde{\lambda}_{D}/\tilde{h}\ll1$
where $\tilde{\lambda}_{D}$ is Debye length. We also assume the presence
of depletion layer of characteristic length $\tilde{\delta}_{D}$,
which is larger than the Debye length (but of the same order of magnitude,
see figure (1b)), $\tilde{\lambda}_{D}/\tilde{h}\ll\tilde{\lambda}_{D}/\tilde{\delta}_{D}<1$,
such that the fluid can be considered Newtonian within the EDL. Furthermore,
we restrict our analysis to shallow flow and negligible inertia 
\begin{equation}
\epsilon=\frac{\tilde{h}}{\tilde{l}}\ll1,\quad\epsilon Re\ll1,\label{Small parameters}
\end{equation}
where the relevant reduced Reynolds number, $\epsilon Re$, is defined in (\ref{eRe}). 

In addition, we assume that surface conduction is negligible, thus
considering a small Dukhin number \citep{lyklemafundamentals},
\begin{equation}
Du=\frac{\tilde{\sigma}_{s}}{\tilde{\sigma}_{b}\tilde{h}}\ll1,\label{Du-1}
\end{equation}
where $\tilde{\sigma}_{s}/\tilde{\sigma}_{b}$ is the ratio of surface to bulk conductivities  \citep[see ][]{lyklemafundamentals}. As noted by \citet{yariv2004electro}
and \citet{khair2008surprising}, this ratio is a length scale which determines
the spatial variations in electric field due to non-uniformity in
surface conduction. High Dukhin numbers would result in significant
spatial variations in electric field, as well as in variations in
concentration which may lead to chemiosmotic flow corrections \citep{derjaguin1961diffusiophoresis,derjaguin1993diffusiophoresis,prieve1984motion,khair2008fundamental}. Under our assumption of
$Du\ll1$, the electric field and the bulk concentration can thus be considered uniform throughout the domain.

Based on the assumptions mentioned above, the relevant governing equations
and boundary conditions are the continuity equation 
\begin{equation}
\tilde{\boldsymbol{\nabla}}\cdot\tilde{\boldsymbol{u}}=0,\label{eq:Continuity}
\end{equation}
the momentum equation
\begin{equation}
\tilde{\rho}\tilde{\boldsymbol{u}}\cdot\boldsymbol{\tilde{\nabla}}\tilde{\boldsymbol{u}}=-\tilde{\boldsymbol{\nabla}}\tilde{p}+\tilde{\boldsymbol{\nabla}}\cdot\boldsymbol{\tilde{\tau}},\label{eq:Momentum}
\end{equation}
the constitutive equation for the stress tensor $\boldsymbol{\tilde{\tau}}$
of power-law fluid
\begin{equation}
\tilde{\boldsymbol{\tau}}=2\tilde{\mu}_{eff}\dot{\gamma}^{n-1}\tilde{\boldsymbol{D}},\label{eq:Stress tensor}
\end{equation}
and the appropriate Helmholtz-Smoluchowski slip boundary conditions \citep{hunter2001foundations} as well as the no-penetration conditions  on the solid walls 
\begin{equation}
\tilde{\boldsymbol{u}}_{\Vert}|_{\tilde{z}=\pm\tilde{h}}=-\frac{\tilde{\varepsilon}\tilde{\zeta}(\tilde{x},\tilde{y})\tilde{\boldsymbol{E}}_{\Vert}}{\tilde{\eta}_{N}},\quad\tilde{\boldsymbol{u}}_{\bot}|_{\tilde{z}=\pm\tilde{h}}=0,\label{H-S slip boundary conditions}
\end{equation}
where $\tilde{\varepsilon}$ is the fluid permittivity and $\tilde{\eta}_{N}$
is the viscosity of Newtonian fluid in the depletion layer. 

The characteristic value of the velocity in the $\tilde{x}-\tilde{y}$
plane, $\tilde{u}^{*},$ is given by the Helmholtz-Smoluchowski slip
condition as $\tilde{u}^{*}=-\tilde{\varepsilon}\tilde{\zeta}^{*}\tilde{E}^{*}/\tilde{\eta}_{N}$,
where $\tilde{\zeta}^{*}$ is characteristic value of zeta potential
and $\tilde{E}{}^{*}$ is characteristic externally applied electric
field. The characteristic velocity in the $\boldsymbol{\hat{z}}$
direction, $\tilde{w}^{*}$, and the characteristic pressure, $\tilde{p}^{*}$,
remain to be determined from scaling arguments. Scaling by the characteristic
dimensions, we define the normalized coordinates, $(x,y,z)=(\tilde{x}/\tilde{l},\tilde{y}/\tilde{l},\tilde{z}/\tilde{h})$,
normalized velocity, $(u,v,w)=(\tilde{u}/\tilde{u}^{*},\tilde{v}/\tilde{u}^{*},\tilde{w}/\tilde{w}^{*})$,
normalized pressure, $p=\tilde{p}/\tilde{p}^{*}$, normalized zeta
potential distribution, $\zeta=\tilde{\zeta}/\tilde{\zeta}^{*}$ and
normalized applied electric field, $\boldsymbol{E_{\Vert}}=\tilde{\boldsymbol{E}}_{\Vert}/\tilde{E}{}^{*}.$

Substituting the normalized parameters into (\ref{eq:Continuity}),
order of magnitude analysis yields
\begin{equation}
\frac{\tilde{w}^{*}}{\tilde{u}^{*}}\sim\frac{\tilde{h}}{\tilde{l}}=\epsilon.\label{eq:Continuity-scaling}
\end{equation}
Substituting (\ref{eq:Continuity-scaling}) into (\ref{eq:Momentum})
and performing order of magnitude analysis, we obtain the characteristic
pressure 
\begin{equation}
\tilde{p}^{*}=\left(\frac{2n+1}{n}\right)^{n}\frac{\tilde{u}^{*n}\tilde{\mu}_{eff}}{\epsilon^{n+1}\tilde{l}^{n}}=\left(\frac{2n+1}{n}\right)^{n}\frac{\tilde{\mu}_{eff}}{\epsilon^{n+1}\tilde{l}^{n}}\left(-\frac{\tilde{\varepsilon}\tilde{\zeta}^{*}\tilde{E}^{*}}{\tilde{\eta}_{N}}\right)^{n},\label{eq:Typical pressure}
\end{equation}
as well as the condition for negligible inertia expressed in terms
of relevant physical quantities,
\begin{equation}
\epsilon Re=\epsilon\frac{\tilde{\rho}\tilde{u}^{*2-n}\tilde{h}^{n}}{\tilde{\mu}_{eff}}=\epsilon\frac{\tilde{\rho}\tilde{h}^{n}}{\tilde{\mu}_{eff}}\left(-\frac{\tilde{\varepsilon}\tilde{\zeta}^{*}\tilde{E}^{*}}{\tilde{\eta}_{N}}\right)^{2-n}\ll1.\label{eRe}
\end{equation}
The expression (\ref{eq:Typical pressure}) is similar, up to constant
factor, to the maximum pressure that can be achieved by electroosmotic
pumps, for the case of strong depletion near the wall \citep{berli2010output}. We also note that the characteristic pressure depends on the ratio
of bulk to depletion layer viscosities, and does not increase linearly
with the applied electric field and in the case of a Newtonian fluid \citep{boyko2015flow}. Using (\ref{eq:Continuity-scaling})
the dimensional shear rate, $\dot{\gamma}$ can be estimated as 
\begin{equation}
\dot{\gamma}=\frac{\tilde{u}^{*}}{\epsilon\tilde{l}}\left(\left|\frac{\partial\boldsymbol{u}_{||}}{\partial z}\right|+O(\epsilon^{2})\right),\label{eq:Approximation  for shear rate}
\end{equation}
where $\left|\partial\boldsymbol{u}_{||}/\partial z\right|=\sqrt{\left(\partial u/\partial z\right)^{2}+\left(\partial v/\partial z\right)^{2}}$. 

Substituting (\ref{eq:Continuity-scaling})-(\ref{eq:Approximation  for shear rate})
into (\ref{eq:Continuity})-(\ref{H-S slip boundary conditions})
and applying the lubrication approximation results in the following
set of normalized equations and boundary conditions, \begin{subequations}
\begin{equation}
\frac{\partial u}{\partial x}+\frac{\partial v}{\partial y}+\frac{\partial w}{\partial z}=0,\label{eq:Continuity Final}
\end{equation}
\begin{equation}
\boldsymbol{\nabla}_{\Vert}p=\left(\frac{n}{2n+1}\right)^{n}\frac{\partial}{\partial z}\left(\left|\frac{\partial\boldsymbol{u}_{||}}{\partial z}\right|^{n-1}\frac{\partial\boldsymbol{u}_{||}}{\partial z}\right)+O(\epsilon Re,\epsilon^{2}),\label{eq:Momentum in-plane-Final}
\end{equation}
\begin{equation}
\frac{\partial p}{\partial z}=O(\epsilon^{3}Re,\epsilon^{2}),\label{eq:Momentum z-Final}
\end{equation}
\begin{equation}
\boldsymbol{u}_{\Vert}|_{z=\pm1}=\zeta(x,y)\boldsymbol{E_{\Vert}},\quad\boldsymbol{u}_{\bot}|_{z=\pm1}=0.\label{Normalized H-S slip boundary conditions}
\end{equation}
\end{subequations}
where $\mathbf{\mathbf{\boldsymbol{\nabla}}_{\Vert}}=\left(\partial/\partial x,\partial/\partial y\right)$ is the two dimensional gradient.

\section{The Governing Equation and its Asymptotic Approximation for Power-Law
Fluids}

Integrating (\ref{eq:Momentum in-plane-Final}) with respect to $z$ and applying the symmetry condition at the mid-plane yields,
\begin{equation}
\boldsymbol{\nabla}_{\Vert}pz=\left(\frac{n}{2n+1}\right)^{n}\left|\frac{\partial\boldsymbol{u}_{||}}{\partial z}\right|^{n-1}\frac{\partial\boldsymbol{u}_{||}}{\partial z}\label{eq:dp*z}
\end{equation}
or alternatively, 
\begin{equation}
\frac{\partial\boldsymbol{u}_{||}}{\partial z}=\frac{2n+1}{n}\left|\boldsymbol{\nabla}_{\Vert}p\right|^{\frac{1}{n}-1}\boldsymbol{\nabla}_{\Vert}p\left|z\right|^{\frac{1}{n}}.\label{eq:du}
\end{equation}
Integrating (\ref{eq:du}) again with respect to $z$ and applying
the slip boundary conditions (\ref{Normalized H-S slip boundary conditions}) we obtain
\begin{equation}
\boldsymbol{u}_{||}=\left(\frac{2n+1}{n+1}\right)\left|\boldsymbol{\nabla}_{\Vert}p\right|^{\frac{1}{n}-1}\boldsymbol{\nabla}_{\Vert}p\left(\left|z\right|^{1+\frac{1}{n}}-1\right)+\zeta(x,y)\boldsymbol{E_{\Vert}}.\label{eq:velocity u}
\end{equation}
Utilizing the continuity equation (\ref{eq:Continuity Final})
and the in-plane velocity (\ref{eq:velocity u}), we obtain
an explicit expression for the perpendicular velocity 
\begin{equation}
\boldsymbol{u}_{\bot}=\left(\frac{n}{n+1}\right)z\left(1-\left|z\right|^{1+\frac{1}{n}}\right)\left[\boldsymbol{E_{\Vert}}\cdot\mathbf{\boldsymbol{\nabla}_{\Vert}}\zeta\right]\boldsymbol{\hat{z}}.\label{Perpendicular velocity}
\end{equation}
Defining the mean in-plane velocity, as $\left\langle \boldsymbol{u}_{\Vert}\right\rangle =(1/2)\intop_{z=-1}^{z=1}\boldsymbol{u}_{\Vert}dz=\intop_{z=0}^{z=1}\boldsymbol{u}_{\Vert}dz,$
and making use of (\ref{eq:Continuity Final}), (\ref{Normalized H-S slip boundary conditions}) and (\ref{eq:velocity u}) yields the depth-averaged equations 
\begin{equation}
\mathbf{\boldsymbol{\nabla}_{\Vert}\cdot}\left\langle \boldsymbol{u}_{\Vert}\right\rangle =0,\label{eq:div of in-plane mean velocity}
\end{equation}
and
\begin{equation}
\left\langle \boldsymbol{u}_{\Vert}\right\rangle =-\left|\boldsymbol{\nabla}_{\Vert}p\right|^{\frac{1}{n}-1}\boldsymbol{\nabla}_{\Vert}p+\zeta(x,y)\boldsymbol{E_{\Vert}}.\label{eq: in-plane mean velocity}
\end{equation}
Applying the two dimensional divergence to (\ref{eq: in-plane mean velocity}),
and using (\ref{eq:div of in-plane mean velocity}), we obtain an
equation in terms of the pressure only,
\begin{equation}
\boldsymbol{\nabla}_{\Vert}\cdot\left(\left|\boldsymbol{\nabla}_{\Vert}p\right|^{\frac{1}{n}-1}\boldsymbol{\nabla}_{\Vert}p\right)=\boldsymbol{E_{\Vert}}\cdot\mathbf{\boldsymbol{\nabla}_{\Vert}}\zeta.\label{eq: pressure only}
\end{equation}
The p-Poisson equation (\ref{eq: pressure only}) describes the pressure
in a Hele-Shaw cell containing a power-law fluid subjected to non-uniform
EOF, and extends the homogenous p-Laplacian equation derived by \citet{aronsson1992hele} for the case of no slip boundary conditions. The source term in (\ref{eq: pressure only}) depends on gradients
of zeta potential which are parallel to the applied electric field,
thus allowing an associated gauge freedom in the
choice of the zeta potential without affecting the resulting pressure. We note that one may eliminate the pressure from (\ref{eq: in-plane mean velocity})
by applying the normal component of the curl operator leading to the
governing equation in terms of the depth-averaged velocity field alone. 

For the case of non-Newtonian power-law fluid that exhibits a weak
shear-thinning or shear-thickening behavior, we define the auxiliary
small parameter $\delta$,
\begin{equation}
n=1-\delta,\label{Asym-n}
\end{equation}
and the asymptotic expansions
\begin{equation}
p=p^{(0)}+\delta p^{(1)}+O(\delta^{2}),\label{Asym p}
\end{equation}
and
\begin{equation}
\left\langle \boldsymbol{u}_{\Vert}\right\rangle =\left\langle \boldsymbol{u}_{\Vert}\right\rangle ^{(0)}+\delta\left\langle \boldsymbol{u}_{\Vert}\right\rangle ^{(1)}+O(\delta^{2}),\label{Asym u}
\end{equation}
where $\left|\delta\right|\ll1$ is a small parameter which is positive
for shear-thinning and negative for shear-thickening behaviors, respectively.
Substituting (\ref{Asym-n}) and (\ref{Asym p}) into the viscous
term in (\ref{eq: pressure only}) and using the expansion $1/n\sim1+\delta+O(\delta^{2})$, yields 
\begin{equation}
\left|\boldsymbol{\nabla}_{\Vert}p\right|^{\frac{1}{n}-1}=1+\delta\ln\left|\boldsymbol{\nabla}_{\Vert}p^{(0)}\right|+O(\delta^{2}).\label{Asym expansion of viscosity}
\end{equation}
Utilizing (\ref{Asym-n})-(\ref{Asym expansion of viscosity}),
the leading order and first order correction of (\ref{eq: in-plane mean velocity}) are
\begin{equation}
O(1):\quad\left\langle \boldsymbol{u}_{\Vert}\right\rangle ^{(0)}=-\boldsymbol{\nabla}_{\Vert}p^{(0)}+\zeta\boldsymbol{E_{\Vert}},\label{Leading order depth averaged velocity}
\end{equation} 
and
\begin{equation}
O(\delta):\quad\left\langle \boldsymbol{u}_{\Vert}\right\rangle ^{(1)}=-\boldsymbol{\nabla}_{\Vert}p^{(1)}-\ln\left|\boldsymbol{\nabla}_{\Vert}p^{(0)}\right|\boldsymbol{\nabla}_{\Vert}p^{(0)},\label{First order depth averaged velocity}
\end{equation}
respectively. Applying the two dimensional divergence to (\ref{Leading order depth averaged velocity})
and (\ref{First order depth averaged velocity}), and using (\ref{eq:div of in-plane mean velocity}),
we obtain equations for the leading order and first order correction of the pressure
\begin{equation}
O(1):\quad\nabla_{\Vert}^{2}p^{(0)}=\boldsymbol{E_{\Vert}}\cdot\mathbf{\boldsymbol{\nabla}_{\Vert}}\zeta,\label{Leading order pressure}
\end{equation}
and
\begin{equation}
O(\delta):\quad\nabla_{\Vert}^{2}p^{(1)}=-\boldsymbol{\nabla}_{\Vert}\cdot\left(\ln\left|\boldsymbol{\nabla}_{\Vert}p^{(0)}\right|\boldsymbol{\nabla}_{\Vert}p^{(0)}\right),\label{First order pressure}
\end{equation}
respectively. Both (\ref{Leading order pressure}) and (\ref{First order pressure})
are Poisson equations where the inhomogeneous part of (\ref{Leading order pressure})
is related to the non-uniform zeta potential distribution \citep[see ][]{boyko2015flow}, and the inhomogeneous part of (\ref{First order pressure})
emanates from the non-Newtonian response of the fluid to the leading
order pressure gradients.

\section{Results - 1D configurations}

We here present a closed form exact solution of (\ref{eq: pressure only})
for arbitrary one dimensional configurations, as well as their asymptotic
approximation. We then use a particular case of an abrupt change in
zeta potential to evaluate the accuracy of the asymptotic approximation. 

\subsection{Exact Solutions of the p-Poisson Equation for 1D configurations}

Consider the case where the zeta potential distribution is a function
of $x$ only, $\zeta=\zeta(x),$ and the electric field is directed
along the $\boldsymbol{\hat{x}}$ axis, $\boldsymbol{E_{\Vert}}=E\boldsymbol{\hat{x}}$.
For such configurations, the problem becomes one dimensional, and
(\ref{eq:velocity u}) and (\ref{eq: pressure only}) take the form
\begin{equation}
q=-\left|\frac{dp}{dx}\right|^{\frac{1}{n}-1}\frac{dp}{dx}+\zeta(x)E,\label{1D flux}
\end{equation}
and
\begin{equation}
\frac{d}{dx}\left(\left|\frac{dp}{dx}\right|^{\frac{1}{n}-1}\frac{dp}{dx}\right)=E\frac{d\zeta(x)}{dx},\label{1D pressure only}
\end{equation}
where $q$ is a volume flux. We prescribe gauge pressure at the inlet
and outlet
\begin{equation}
p|_{x=0}=p|_{x=1}=0,\label{1D BC}
\end{equation}
or alternatively impose inlet pressure $p|_{x=0}=0$ and a desired
volume flux $q$. Solving (\ref{1D pressure only}), we obtain
\begin{equation}
p(x)=\int_{0}^{x}\left|\zeta(x)E-q\right|^{n-1}\left(\zeta(x)E-q\right)dx,\label{1D pressure distribution}
\end{equation}
where $q$ may be calculated (in the case of (\ref{1D BC})) from
the equality 
\begin{equation}
\int_{0}^{1}\left|\zeta(x)E-q\right|^{n-1}\left(\zeta(x)E-q\right)dx=0.\label{1D flux calculation}
\end{equation}
We consider a case where the zeta potential acquires a constant positive
value, $\zeta_{0}$, before the discontinuity at $x_{0}$ and vanishes afterwards, 
\begin{equation}
\zeta(x)=\zeta_{0}H(x_{0}-x),\label{eq:Non-uniform zeta 1D}
\end{equation}
where $H$ is the Heaviside function. 
Substituting (\ref{eq:Non-uniform zeta 1D}) into (\ref{1D flux calculation}) yields 
\begin{equation}
\int_{0}^{x_{0}}\left|\zeta_{0}E-q\right|^{n-1}\left(\zeta_{0}E-q\right)dx-\int_{x_{0}}^{1}\left|q\right|^{n-1}qdx=0,\label{eq: governing equation for pressure, Heaviside H-S}
\end{equation}
Assuming that $E$ is positive and $E\zeta_{0}\geq q$ (since the value
$E\zeta_{0}$ is the maximal volume flux that can be achieved, when
entire channel has a zeta potential $\zeta_{0})$, from (\ref{eq: governing equation for pressure, Heaviside H-S})
we obtain
\begin{equation}
\frac{q}{E\zeta_{0}}=\frac{1}{1+\left((1-x_{0})/x_{0}\right)^{1/n}}.\label{velocity distribution 1D}
\end{equation}
Substituting (\ref{velocity distribution 1D}) into (\ref{1D pressure distribution}) results in
\begin{equation}
p(x)=\left\{\begin{array}{ll}
\displaystyle{-q^{n}\left(x-1\right)} & x>x_{0}\vspace{1mm}\\
\displaystyle{\left(\zeta_{0}E-q\right)^{n}x} & x\leqslant x_{0}.
\end{array}\label{eq: pressure, Heaviside H-S+BC} \right.
\end{equation}

\begin{figure}
\centerline{\includegraphics[scale=0.48]{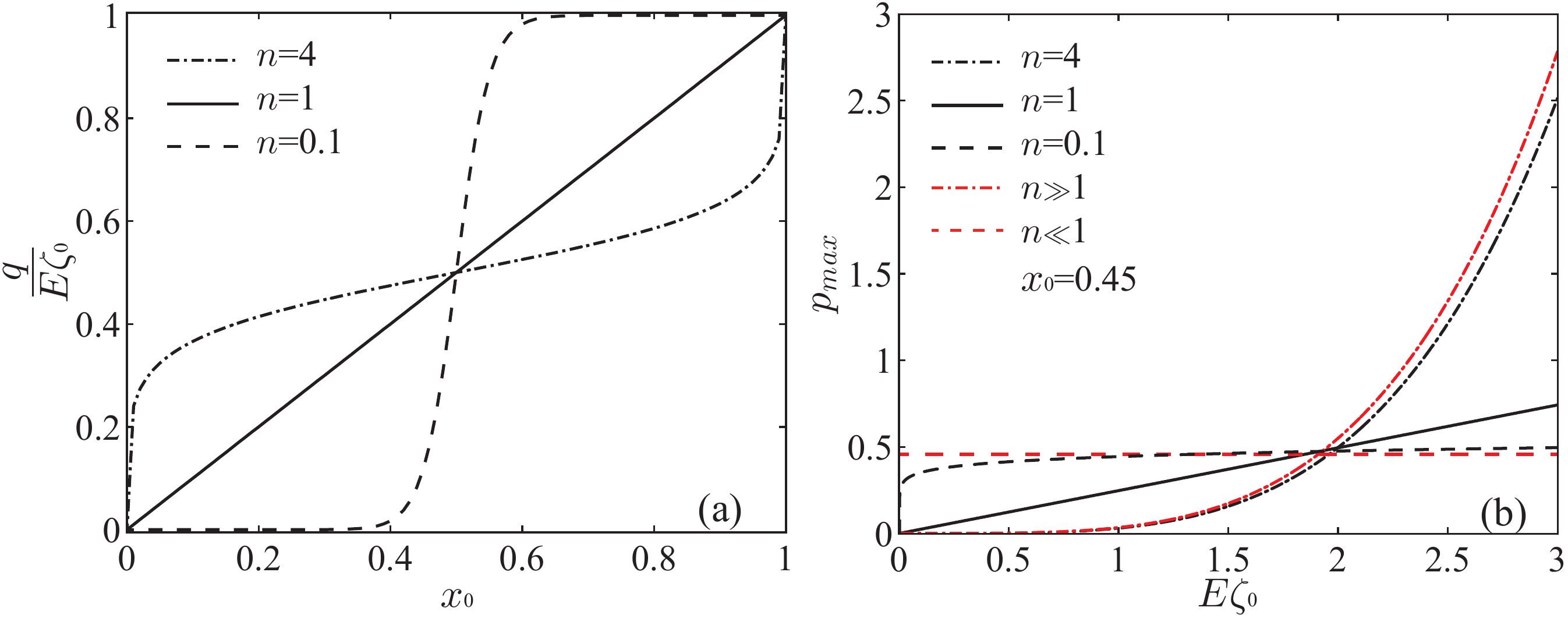}}
\caption{The volume flux and maximal pressure in the case of a step zeta potential
distribution for one dimensional configurations. (a) Comparison of
flux to slip velocity ratio, $q/E\zeta_{0}$, as function of the step
location $x_{0}$ for strong shear-thickening (dashed-dot line), strong
shear-thinning (dashed line) and Newtonian behavior. (b) Maximal pressure
obtained from (\ref{Maximal pressure}) as function of $E\zeta_{0}$
for $x_{0}=0.45$ and $n=0.1,1,4$. Black dashed line describes a
strong shear-thinning behavior which tends to constant value given
by (\ref{Maximal pressure small n}) and shown as red dashed line.
Black dashed-dot line corresponds to the case of strong shear-thickening,
and coincides with asymptotic behavior given by (\ref{Maximal pressure large n})
(red dashed-dot line). The black solid line presents a Newtonian case.}
\end{figure}

Figure (2a) presents the volume flux to slip velocity ratio, $q/E\zeta_{0}$,
as a function of $x_{0}$ (the position of the zeta potential discontinuity) for a strongly shear-thinning fluid $n=0.1$
(dashed line), a strongly shear-thickening fluid $n=4$ (dashed-dotted
line), and a Newtonian fluid (solid line). For a Newtonian fluid and
constant $E\zeta_{0}$, the flux $q$ increases linearly with $x_{0}$,
as expected. In the case of strong shear-thinning, the flux to slip
velocity ratio, $q/E\zeta_{0}$, is approximately zero when the abrupt
change in zeta potential occurs before the middle of the channel ($x_{0}<0.5$),
while $q/E\zeta_{0}$ is approximately 1 in case the discontinuity
is positioned after the middle of the channel ($x_{0}>0.5$). In contrast,
in the case of strong shear-thickening, $q/E\zeta_{0}$ is weakly
dependent on the location of the discontinuity (except near the boundaries). 

It is also interesting to investigate the effect of the parameter
$n$ on resulting maximal pressure in the channel, given by 
\begin{equation}
p_{max}\triangleq p_{max}^{ex}=\left(E\zeta_{0}\right)^{n}\frac{1-x_{0}}{\left(1+\left((1-x_{0})/x_{0}\right)^{1/n}\right)^{n}}.\label{Maximal pressure}
\end{equation}
When $n\ll1$, we have 
\begin{equation}
p_{max}\sim\left\{\begin{array}{ll}
1-x_{0} & x_{0}>0.5\\
x_{0} & x_{0}<0.5
\end{array},\label{Maximal pressure small n} \right.
\end{equation}
independently of value $E\zeta_{0}$, whereas for $n\gg1$, we obtain
\begin{equation}
p_{max}\sim\left(1-x_{0}\right)\left(\frac{1}{2}E\zeta_{0}\right)^{n}.\label{Maximal pressure large n}
\end{equation}
Figure (2b) presents a variation of maximal pressure as function of
$E\zeta_{0}$ for a specific case of $x_{0}=0.45.$ As expected, for
a Newtonian fluid $\left(n=1\right)$ we obtain a linear behavior
(black solid line). For small values of $n$, the maximal pressure
(black dashed lines) rapidly approaches a constant value given by
(\ref{Maximal pressure small n}), which is solely determined by the
location of discontinuity $x_{0}$ (red dashed line). On the other
hand, for large values of $n$ the maximal pressure (black dashed-dot
lines) approaches the asymptotic limit (\ref{Maximal pressure large n})
(red dashed-dot line).We note that for sufficiently small values of
$E\zeta_{0}$, corresponding to shear rate less than one, where the
viscosity of shear-thinning fluid exceeds the viscosity of the shear-thickening
fluid, resulting in larger maximal pressure of shear-thinning fluid
in this region. 

\subsection{Asymptotic approximation for 1D configurations }

The leading and first order asymptotic approximations for one dimensional
configurations are\begin{subequations}
\begin{equation}
O(1):\quad\frac{d^{2}p^{(0)}}{dx^{2}}=E\frac{d\zeta}{dx},\label{Leading order pressure 1D}
\end{equation}
and
\begin{equation}
O(\delta):\quad\frac{d^{2}p^{(1)}}{dx^{2}}=-\frac{d}{dx}\left(\ln\left|\frac{dp^{(0)}}{dx}\right|\frac{dp^{(0)}}{dx}\right).\label{First order pressure 1D}
\end{equation}
\end{subequations} (\ref{Leading order pressure 1D}) and (\ref{First order pressure 1D})
are readily solved, to attain\begin{subequations}
\begin{equation}
p^{(0)}(x)=\int_{0}^{x}E\zeta(x)dx-x\int_{0}^{1}E\zeta(x)dx.\label{Leading order pressure general solution 1D}
\end{equation}
and
\begin{equation}
p^{(1)}(x)=-\int_{0}^{x}\ln\left|\frac{dp^{(0)}}{dx}\right|\frac{dp^{(0)}}{dx}dx+x\int_{0}^{1}\ln\left|\frac{dp^{(0)}}{dx}\right|\frac{dp^{(0)}}{dx}dx.\label{First order pressure general solution}
\end{equation}
\end{subequations}We now derive the asymptotic solution corresponding
to a step function in zeta potential, (\ref{eq:Non-uniform zeta 1D}),
in order to compare the exact and asymptotic results and evaluate
the accuracy of the asymptotic solution. We define $x_{0}=0.5$. and
substitute (\ref{eq:Non-uniform zeta 1D}) into (\ref{Leading order pressure 1D First example-1}), yielding
\begin{equation}
\frac{d^{2}p^{(0)}}{dx^{2}}=-2E\zeta_{0}\delta(1-2x),\label{Leading order pressure 1D First example-1}
\end{equation}
and the corresponding solution of (\ref{Leading order pressure 1D First example-1}) is 
\begin{equation}
p^{(0)}(x)=\frac{1}{2}E\zeta_{0}\left(x+\left(1-2x\right)H(2x-1)\right).\label{Leading order pressure 1D First example solution-1}
\end{equation}
Substituting (\ref{Leading order pressure 1D First example solution-1})
into (\ref{First order pressure general solution}) yield 
\begin{equation}
\frac{d^{2}p^{(1)}}{dx^{2}}=2E\zeta_{0}\ln\left(\frac{1}{2}E\zeta_{0}\right)\delta(1-2x).\label{First order pressure 1D First example NN-1}
\end{equation}

\begin{figure}
\centerline{\includegraphics[scale=0.45]{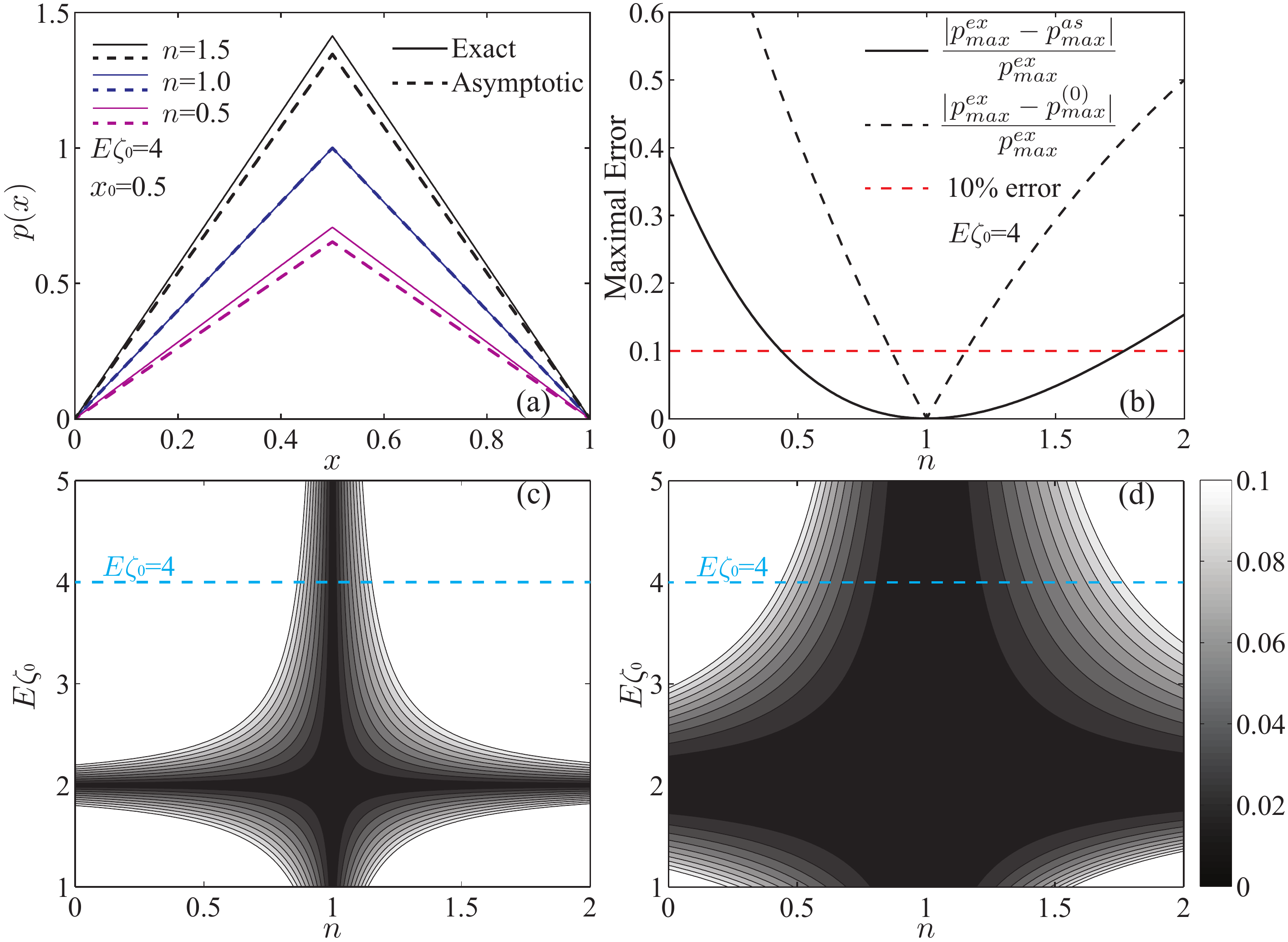}}
\caption{Comparison between exact solution and asymptotic approximations in
the case of a step zeta potential distribution in one dimension. (a)
The resulting pressure distribution calculated from exact (solid lines)
and asymptotic (dashed lines) solutions for $n=0.5,1,1.5$, $x_{0}=0.5$
and $E\zeta_{0}=4$. (b) The error in maximal pressure between exact
and asymptotic solutions as a function of $n$, for the case of $x_{0}=0.5$
and $E\zeta_{0}=4$. Black dashed line given by (\ref{Error 0}) corresponds
to the Newtonian approximation, while black solid line corresponds
to the first order correction (\ref{Error 0+1}). Red dashed line
represents a $10\,\%$ error. (c,d) Colormaps of the maximal error
as function of the parameter $n$ and slip velocity $E\zeta_{0}$,
for the cases of a Newtonian approximation and a first order asymptotic
approximation, respectively. The cyan dashed line shown in (c) and
(d) corresponds to $E\zeta_{0}=4$, shown in (b). As expected, range
of values of $n$ for which error is less than $10\,\%$ is significantly
increased when the first order non-Newtonian correction included.}
\end{figure}

Solving (\ref{First order pressure 1D First example NN-1}) before
and after the discontinuity, following the same procedure as we illustrated
above, and requiring the continuity of the pressure and of the flux at
$x_{0}=0.5$, leads to first order correction for the pressure distribution
\begin{equation}
p^{(1)}(x)=-\frac{1}{2}E\zeta_{0}\ln\left(\frac{1}{2}E\zeta_{0}\right)\left(x+\left(1-2x\right)H(2x-1)\right).\label{First order pressure 1D First example NN solution-1}
\end{equation}
Combining (\ref{Leading order pressure 1D First example solution-1})
and (\ref{First order pressure general solution}), and substituting
$x=x_{0}=0.5$ yield a maximal pressure as a function of $E\zeta_{0}$
and $\delta$
\begin{equation}
p_{max}^{as}\triangleq\frac{1}{4}E\zeta_{0}\left(1-\delta\ln\left(\frac{1}{2}E\zeta_{0}\right)\right).\label{1D First example maximal pressure as}
\end{equation}
Figure (3a) presents the pressure distribution obtained from exact
(solid lines) and asymptotic (dashed lines) solutions for $n=0.5,1,1.5$,
$x_{0}=0.5$ and $E\zeta_{0}=4$. Since (\ref{eq: pressure, Heaviside H-S+BC})
is an exact solution, we define the maximal error of the leading order
and first order asymptotic solutions \begin{subequations}
\begin{equation}
e^{(0)}=\left|\frac{p_{max}^{ex}-p_{max}^{(0)}}{p_{max}^{ex}}\right|,\label{Error 0}
\end{equation}
and
\begin{equation}
e^{(0)+(1)}=\left|\frac{p_{max}^{ex}-p_{max}^{as}}{p_{max}^{ex}}\right|,\label{Error 0+1}
\end{equation}
\end{subequations}respectively. 
Figure (3b) presents the maximal error between the exact solution of the
p-Poisson equation and the asymptotic approximation for a range of
$n$ values, for the case of $x_{0}=0.5$ and $E\zeta_{0}=4$. 
The solid black line presents the error of the asymptotic solution
which includes the first order correction, whereas the black dashed
line shows the maximal error given by (\ref{Error 0}) in the case
of a Newtonian approximation (zeroth order only). The red dashed line
presents a $10\,\%$ error. Clearly, the range of values of $n$ for
which the error is less than $10\,\%$ is significantly increased
when the first order correction is taken into account.

Figure (3c) and (3d) show colormaps of the maximal error as function
of parameter $n$ and slip velocity $E\zeta_{0}$, in the case Newtonian
approximation and non-Newtonian correction, respectively. The cyan
dashed line in figures (3c) and (3d) represents the value of $E\zeta_{0}=4$,
which is presented in figure (3b). As can be inferred from the results of figure (3), the first order
correction significantly reduces the error and also remains applicable
for a wide range of $n$ values. This range of asymptotic applicability
reduces however as $E\zeta_{0}$ increases.

\section{Results - 2D configurations }

\subsection{Axi-symmetric pressure fields}

Assuming a uniform electric field along the $\boldsymbol{\hat{x}}$
axis, $\boldsymbol{E_{\Vert}}=E\boldsymbol{\hat{x}}$, we consider
the case where the gradient of zeta potential in $\boldsymbol{\hat{x}}$
direction, $\partial\zeta/\partial x$, is function of a single argument
$r=\sqrt{x^{2}+y^{2}}$. Therefore, the resulting pressure also depends
only on $r$ and is obtained through 
\begin{equation}
\frac{1}{r}\frac{d}{dr}\left(\left|\frac{dp}{dr}\right|^{\frac{1}{n}-1}\frac{dp}{dr}r\right)=E\frac{\partial\zeta}{\partial x}(r).\label{2Dradial pressure}
\end{equation}
Equation (\ref{2Dradial pressure}) is subjected to a symmetry boundary
condition
\begin{equation}
\frac{dp}{dr}=0\quad\mathrm{at}\quad r=0,\label{Symmetry of dp/dr at r=00003D0-1}
\end{equation}
and the pressure is required to decay far from the actuation region,
\begin{equation}
p=0\quad\mathrm{as}\quad r\rightarrow\infty.\label{Far condition for pressure}
\end{equation}
Integrating (\ref{2Dradial pressure}) with respect to $r$, we obtain 
\begin{equation}
\left|\frac{dp}{dr}\right|^{\frac{1}{n}-1}\frac{dp}{dr}=\frac{E}{r}\left(\int r\frac{\partial\zeta}{\partial x}(r)dr+A\right),\label{2Dradial pressure 2}
\end{equation}
where $A$ is a constant determined using (\ref{Symmetry of dp/dr at r=00003D0-1}) 
\begin{equation}
A=-\lim_{r\to0}\frac{1}{r}\int r\frac{\partial\zeta}{\partial x}(r)dr.\label{Limit condition}
\end{equation}
From (\ref{eq: in-plane mean velocity}) and (\ref{2Dradial pressure 2}),
it follows that the resulting flow field in this case is two dimensional,
and identical for Newtonian and non-Newtonian fluids independently of the parameter $n$.

\begin{figure}
\centerline{\includegraphics[scale=0.55]{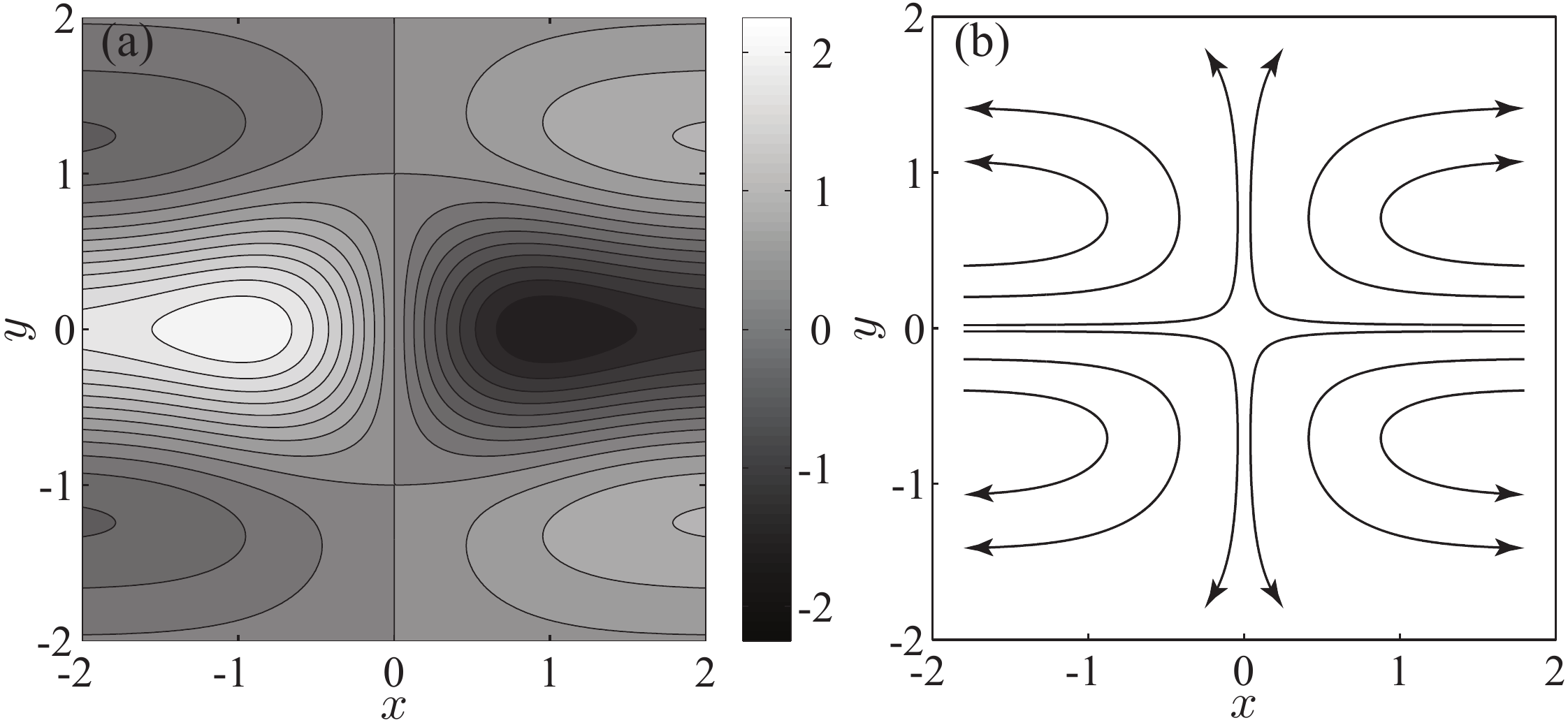}}
\caption{(a) Illustration of two dimensional zeta potential distribution with
an axi-symmetric gradient in the $\boldsymbol{\hat{x}}$ direction,
described by (\ref{zeta(x,y)}). (b) The resulting two dimensional
flow field, due to zeta potential distribution shown in (a). The streamlines
describe stagnation-like flow approaching the origin from the both
side of $\boldsymbol{\hat{x}}$ axis. The flow field is identical
for both Newtonian and non-Newtonian fluids as given by (\ref{2D velocity}).
All calculations were performed using $E=3$ and $\zeta_{0}=\alpha=1$.}
\end{figure}

\begin{table}
  \begin{center}
\def~{\hphantom{0}}
  \begin{tabular}{lccc}
      ~~~~Polymer solution  &  Background electrolyte (solvent)   &   $n$ & $\tilde{\mu}_{eff}$  [Pa\,s$^{n}$]\\[3pt]
\hline
       Polyacrylic acid (PAA),  & 10 mM Tris, & ~~0.38~ & 0.38\\
        ~~8.33 mM, 42 kD, pH=8.2~   & 5 mM acetic acid &\\
       \hline
       Carboxymethyl cellulose (CMC),  & 15 mM phosphoric acid & ~~0.51~ & 1.75\\
        ~~~~~~~1 \% w/v, pH=7 ~   & ~~~\\
      \end{tabular}
  \caption{Rheological parameters of polymer solutions at 25 $^{0}$C. The background aqueous solution has viscosity of $\eta_{N}$=1 mPa\,s. PAA parameters are adapted from \citet{paul2008electrokinetic} and \citet{berli2010output}. CMC experimental data is taken from \citet{olivares2009eof}.}
  \label{tab:kd}
  \end{center}
\end{table}
Hereafter, we focus on configurations where $\zeta(r)$ is continuous at the origin, and thus $A=0$. Integrating
once again (\ref{2Dradial pressure 2}) and using (\ref{Far condition for pressure}),
yields a closed form expression for the pressure
\begin{equation}
p(r)=E^{n}\int_{\infty}^{r}\frac{1}{r^{n}}\left|\int r\frac{\partial\zeta}{\partial x}(r)dr\right|^{n-1}\left(\int r\frac{\partial\zeta}{\partial x}(r)dr\right)dr.\label{Closed form solution-2D pressure}
\end{equation}
For a Newtonian fluid, a Gaussian-like pressure
\begin{equation}
p=E\zeta_{0}\exp\left(-\frac{x^{2}+y^{2}}{\alpha^{2}}\right)=E\zeta_{0}\exp\left(-\frac{r^{2}}{\alpha^{2}}\right),\label{Gaussian}
\end{equation}
where $\alpha$ and $\zeta_{0}$ are positive constants, may be obtained
using the following zeta distribution
\begin{equation}
\zeta(x,y)=-\frac{\zeta_{0}}{\alpha^{4}}\left(2x\alpha+\sqrt{\pi}\left(-2y^{2}+\alpha^{2}\right)\exp\left(\frac{x^{2}}{\alpha^{2}}\right)\mathrm{erf}\left(\frac{x}{\alpha}\right)\right)\exp\left(-\frac{x^{2}+y^{2}}{\alpha^{2}}\right),\label{zeta(x,y)}
\end{equation}
which is shown in figure (4a). We note that in this case the limit
(\ref{Limit condition}) gives $A=0$. 

We here examine the effect of non-Newtonian behavior on the pressure
distribution, resulting from the same zeta potential (\ref{zeta(x,y)}).
Substituting (\ref{zeta(x,y)}) into (\ref{Closed form solution-2D pressure})
provides a closed form analytical solution for the pressure for a non-Newtonian fluid with arbitrary $n$
\begin{equation}
p(r)=n^{-\frac{1}{2}\left(n+1\right)}\left(\frac{\alpha}{2}\right)^{1-n}\left(E\zeta_{0}\right)^{n}\Gamma\left(\frac{n+1}{2};\frac{nr^{2}}{\alpha^{2}}\right),\label{Solution for the pressure 2D}
\end{equation}
where $\Gamma(a;x)=\int_{x}^{\infty}\exp\left(-t\right)t^{a-1}dt$
is the incomplete Gamma function \citep{abramowitz1964handbook}. The
leading order of the asymptotic solution is simply given in (\ref{Gaussian}), while the first order is
\begin{equation}
p^{(1)}(r)=\frac{1}{2}E\zeta_{0}\exp\left(-\frac{r^{2}}{\alpha^{2}}\right)\left(2+\exp\left(\frac{r^{2}}{\alpha^{2}}\right)\mathrm{Ei}\left(-\frac{r^{2}}{\alpha^{2}}\right)-2\ln\left(\frac{2E\zeta_{0}}{\alpha^{2}}r\exp\left(-\frac{r^{2}}{\alpha^{2}}\right)\right)\right),\label{Solution for the pressure 2D-FO}
\end{equation}
where $\mathrm{Ei}(x)=-\int_{-x}^{\infty}\exp\left(-t\right)/tdt$
is the exponential integral function \citep{abramowitz1964handbook}.
The corresponding flow field is readily obtained by differentiating (\ref{Gaussian}) and using (\ref{zeta(x,y)})
\begin{equation}
\left\langle \boldsymbol{u}_{\Vert}\right\rangle =\frac{E\zeta_{0}}{\alpha^{2}}\exp\left(-\frac{x^{2}+y^{2}}{\alpha^{2}}\right)\left(\begin{array}{c}
\left(2\left(1-\frac{1}{\alpha}\right)x-\frac{\sqrt{\pi}}{\alpha^{2}}\left(-2y^{2}+\alpha^{2}\right)\exp\left(\frac{x^{2}}{\alpha^{2}}\right)\mathrm{erf}\left(\frac{x}{\alpha}\right)\right)\boldsymbol{\hat{x}}\\
2y\boldsymbol{\hat{y}}
\end{array}\right).\label{2D velocity}
\end{equation}
Figure (4b) presents several streamlines given by (\ref{2D velocity})
for $E\zeta_{0}=3$. The streamlines describe an incoming fluid flowing
along the both sides $\boldsymbol{\hat{x}}$ axis toward the origin,
followed by outgoing flow initially turning in $\pm\boldsymbol{\hat{y}}$
directions and then curving back in the $\pm\boldsymbol{\hat{x}}$
directions to satisfy mass conservation. The resulting flow field
(\ref{2D velocity}) is independent of parameter $n$, due to a Neumann-type
symmetry boundary condition (\ref{Symmetry of dp/dr at r=00003D0-1})
at the origin. 
\begin{figure}
\centerline{\includegraphics[scale=0.5]{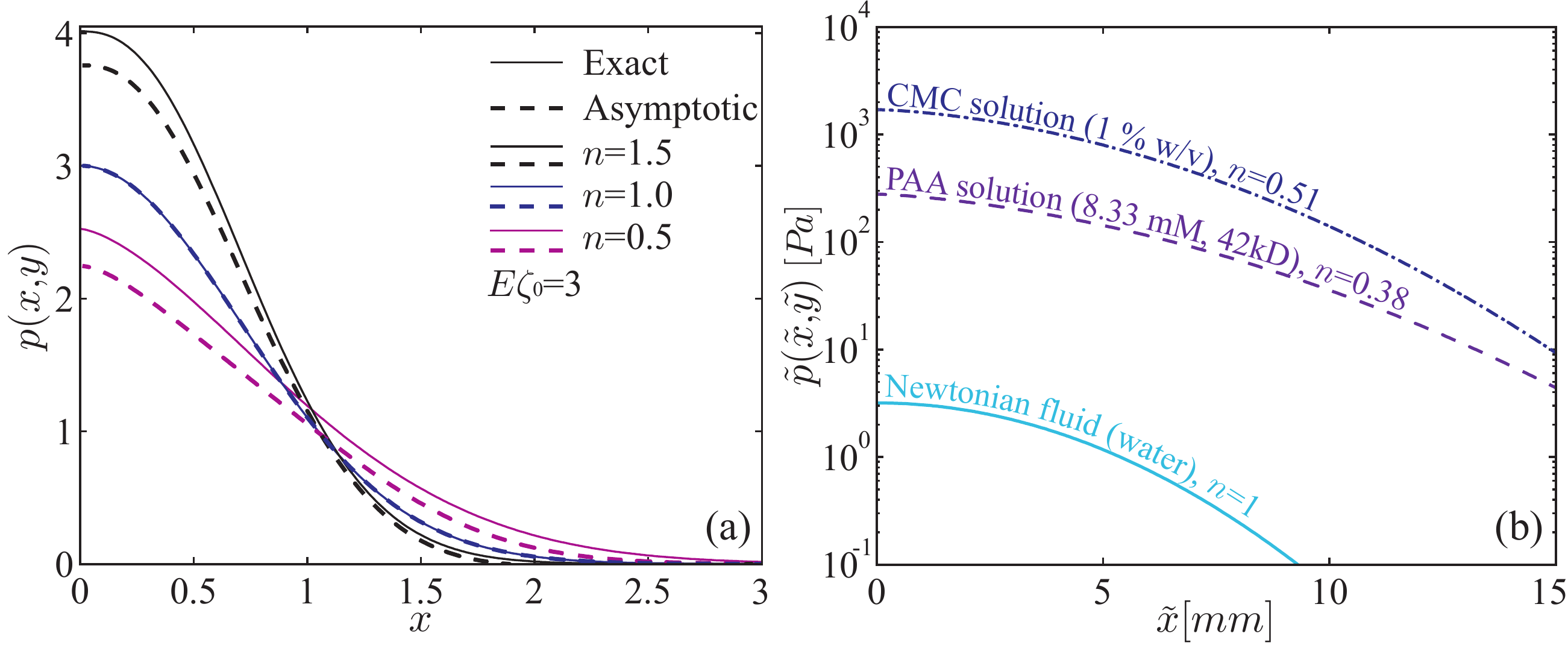}}
\caption{The use of non-Newtonian fluids allows to significantly increase the
pressure resulting from non-uniform EOF by more than two order of
magnitude. (a) Comparison of exact results (solid lines) and asymptotic
approximations (dashed lines) for axi-symmetric (non-dimensional)
pressure distributions, resulting from a two dimensional zeta potential,
(\ref{zeta(x,y)}), for $n=0.5,1,1.5$ and $E\zeta_{0}=3$. (b) Exact
solutions in dimensional form showing significantly greater pressures
for EOF driven flow of non-Newtonian fluids compared with Newtonian
fluids. The different lines correspond to a Newtonian fluid (cyan
solid line), 8.33 mM polyacrilic acid (PAA) (purple dashed line) and
$1\,\%$ carboxymethyl cellulose (CMC) (blue dashed-dot line) solutions.
The rheological power-law properties of PAA and CMC solutions are
available in table 1. For a zeta potential difference of 75 mV, and
an applied electric field of 100 $\mathrm{V\,cm^{-1}}$, a Newtonian
fluid yields a maximum pressure of 3 Pa, whereas as PAA and CMC solutions
yield pressures of 280 and 1700 Pa, respectively. All calculations
were performed using $\tilde{l}=5$ mm, $\tilde{h}=50\,\mu m$, $\tilde{\zeta}^{*}=-25$
mV, $\tilde{E}^{*}=$100 $\mathrm{V\,cm^{-1}}$, $\tilde{\varepsilon}=7.08\cdot10^{-10}\,\mathrm{F\,m^{-1}}$,
$E\zeta_{0}=3$ and $\alpha=1$.}
\end{figure}

In figure (5a) we compare the resulting exact (solid lines) and asymptotic
approximations (dashed lines) for the pressure distribution given
by (\ref{Solution for the pressure 2D}), (\ref{Gaussian}) and (\ref{Solution for the pressure 2D-FO}),
showing good agreement. Using (\ref{Error 0})
and (\ref{Error 0+1}), we also calculated the maximal errors in this
case, concluding that in two-dimensions the accuracy of the asymptotic
solution is approximately the same as in one dimension, yielding error
below $10\,\%$ for values of $n$ lying between 0.5 and 1.75.

Figure (5b) presents the resulting dimensional pressure distribution
of (\ref{Solution for the pressure 2D}), where we used three experimental
values of $n$ and $\tilde{\mu}_{eff}$ from table 1, corresponding
to a Newtonian fluid (cyan solid line), 8.33 mM polyacrilic acid (PAA)
(purple dashed line) and $1\,\%$ carboxymethyl cellulose (CMC) (blue
dashed-dot line) solutions. As can be observed, by leveraging non-Newtonian
polymer solutions, the resulting pressure is increased by two to three
orders of magnitude as compared to a Newtonian fluid. 

\subsection{Asymptotic approximation for a circular spot with constant zeta potential}

We here study the effect of a spot with non-zero zeta potential on the resulting pressure and flow fields. This simple, yet fully two-dimensional case, is useful in providing physical insight on the effect of the power-law fluid on the flow field. In the following, we adopt the superscripts $in$ and $out$
to distinguish between physical quantities in each one of the two
regions. The corresponding zeta potential distribution is given by
\begin{equation}
\zeta(r)=\zeta_{0}H(r_{0}-r).\label{RadialSymDistribution}
\end{equation}
Here $H$ stands for the Heaviside function and $r_{0}$ is the radius
of the spot. Without loss of generality, we assume that the uniform
electric field is directed along the $\boldsymbol{\hat{x}}$ axis,
$\boldsymbol{E_{\Vert}}=E\boldsymbol{\hat{x}}$. Substituting (\ref{RadialSymDistribution})
into (\ref{Leading order pressure}) and taking $p^{(0)}\rightarrow0$
as $r\rightarrow\infty$, yield the leading order solution \citep{boyko2015flow}
\begin{equation}
p^{(0)}(r,\theta)=\left\{\begin{array}{ll}
\displaystyle{\frac{3E\zeta_{0}r_{0}^{2}}{2r}\cos(\theta)} \hspace{0.1in} & r>r_{0}\\
\displaystyle{\frac{3}{2}E\zeta_{0}r\cos(\theta)} & r<r_{0},
\end{array}\label{General solution LO p(r,th) dipole}\right.
\end{equation}
with corresponding flow field
\begin{equation}
\left\langle \boldsymbol{u}_{\Vert}\right\rangle ^{(0)}=\left\{\begin{array}{ll}
\displaystyle{\frac{1}{2}\frac{E\zeta_{0}r_{0}^{2}}{r^{2}}\left(\cos(\theta)\boldsymbol{\hat{r}}+\mathrm{sin}(\theta)\boldsymbol{\hat{\theta}}\right)}=\frac{1}{2}\frac{E\zeta_{0}r_{0}^{2}}{(x^{2}+y^{2})^{2}}\left(\left(x^{2}-y^{2}\right)\boldsymbol{\hat{x}}+2xy\boldsymbol{\hat{y}}\right)\hspace{0.1in} & r>r_{0}\\
\displaystyle{\frac{1}{2}E\zeta_{0}\left(\cos(\theta)\boldsymbol{\hat{r}}-\mathrm{sin}(\theta)\boldsymbol{\hat{\theta}}\right)}=\frac{1}{2}E\zeta_{0}\boldsymbol{\hat{x}}\hspace{0.1in} & r<r_{0}.
\end{array}\label{LO velocity dipole}\right.
\end{equation}
Next, we turn to solve the non-Newtonian (first order) contribution
to the pressure. The source term in the first order equation (\ref{First order pressure})
can be expressed using (\ref{General solution LO p(r,th) dipole}) as
\begin{equation}
-\boldsymbol{\nabla}_{\Vert}\cdot\left(\left(\ln\left|\boldsymbol{\nabla}_{\Vert}p^{(0)}\right|\right)\boldsymbol{\nabla}_{\Vert}p^{(0)}\right)=-\frac{E\zeta_{0}r_{0}^{2}}{r^{3}}H\left(r-r_{0}\right)\cos(\theta).\label{The source term in FO}
\end{equation}
Assuming the
form $p^{(1)}(r,\theta)=f^{(1)}(r)\cos(\theta),$ we obtain an ordinary
differential equation for $f^{(1)}(r)$
\begin{equation}
\frac{1}{r}\frac{d}{dr}\left(r\frac{df^{(1)}(r)}{dr}\right)-\frac{f^{(1)}(r)}{r^{2}}=-\frac{E\zeta_{0}r_{0}^{2}}{r^{3}}H\left(r-r_{0}\right).\label{FO ODE for f(r) NN diploe}
\end{equation}
Solving (\ref{FO ODE for f(r) NN diploe}) and requiring regularity, leads to 
\begin{equation}
p^{(1)}(r,\theta)=\left\{\begin{array}{ll}
\displaystyle{\left(\frac{a^{out}}{r}+\left(b^{out}-\frac{1}{4}E\zeta_{0}\right)r-\frac{1}{4}\frac{E\zeta_{0}r_{0}^{2}}{r}\left(\ln\left(\frac{r_{0}}{r}\right)^{2}-1\right)\right)\cos(\theta)} & r>r_{0}\\
b^{in}r\cos(\theta) & r<r_{0},
\end{array}\label{General solution FO NN p(r,th) dipole} \right.
\end{equation}
where $a^{out}$, $b^{in}$ and $b^{out}$ are coefficients to be
determined from boundary conditions. Demanding vanishing pressure far from the disk, we find
\begin{figure}
\centerline{\includegraphics[scale=0.5]{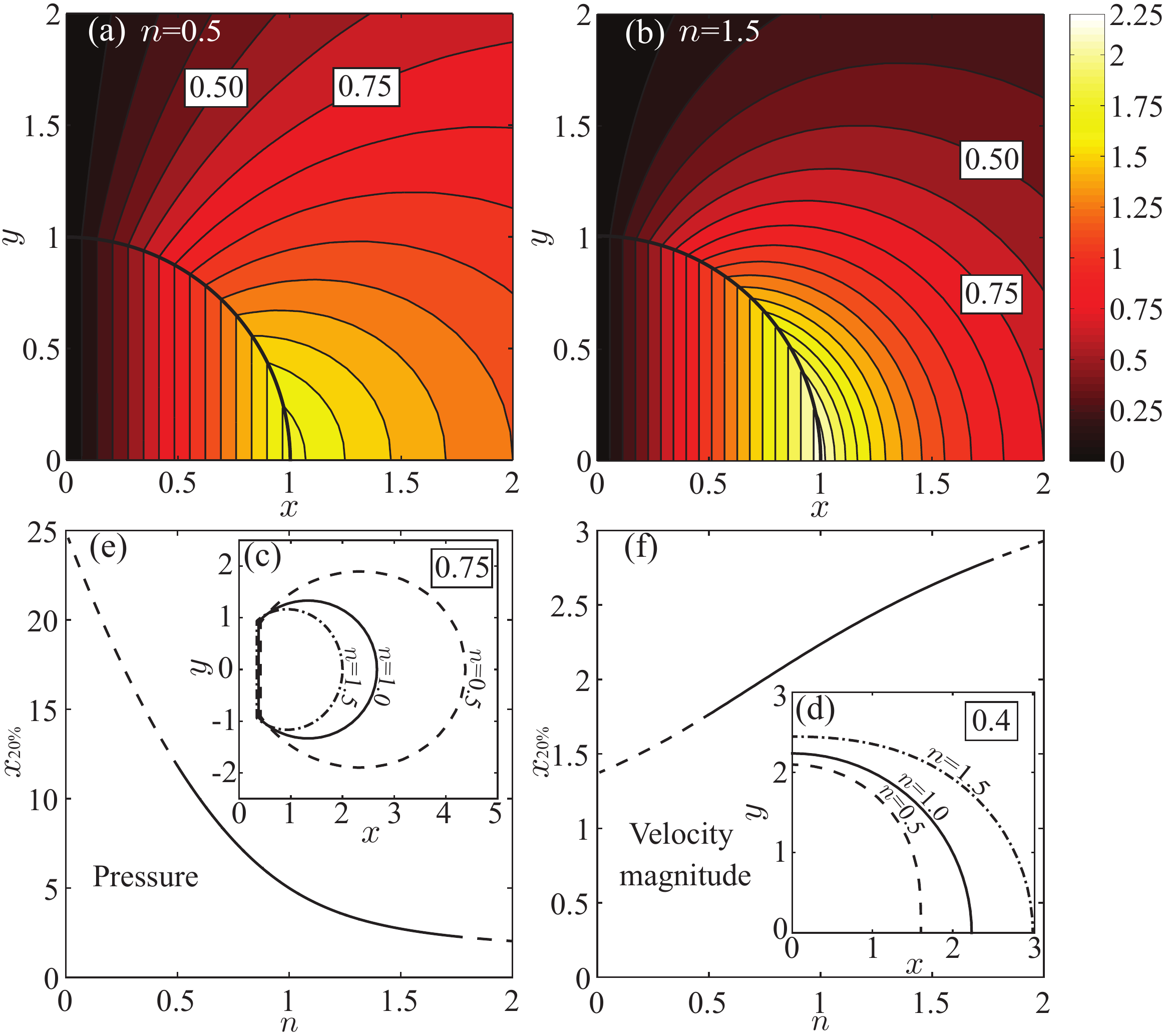}}
\caption{Two dimensional asymptotic approximations for the velocity and pressure
fields of power-law fluids in the presence of a spot with constant
zeta potential. (a) and (b) The pressure distribution (color-map)
in first quadrant for $n=0.5$ and $n=1.5$ corresponding to shear-thinning
and shear-thickening behavior, respectively, showing that the isobars
of shear-thinning fluid are more spaced compared with those of a shear-thickening
fluid. Isobars $p^{as}=0.75$ presented in (c) and contours of velocity
magnitude $\left|\left\langle \boldsymbol{u}_{\Vert}\right\rangle ^{as}\right|=0.4$
presented in (d) correspond to $n=0.5$ (black dashed line), $n=1$
(black solid line) and $n=1.5$ (black dashed-dot line). Note that
contours in (d) are more closely spaced than the contours in (c),
thus indicating weakly dependence on $n$ for the shear-thickening
fluid. (e) and (f) The distance from the origin along the $\boldsymbol{\hat{x}}$
axis, where the pressure and the velocity magnitude drops to $20\%$
of its maximum value, as function of $n$, respectively. Solid part
of the line in (e) and (f) corresponds to the range of $n$ lying
between 0.5 and 1.75, where the asymptotic approximation is assumed
to be valid, while the dashed line corresponds region that may involve
significant errors. It follows from (c)-(f) that the pressure distribution
strongly dependent on $n$ for $n<1$ and weakly dependent on $n$
for $n>1$ as compared to velocity magnitude showing only weak dependence
$n$. All calculations were performed using $E\zeta_{0}=4$ and $r_{0}=1$.}
\end{figure}
\begin{equation}
b^{out}=\frac{1}{4}E\zeta_{0}.\label{bout coefficient}
\end{equation}
The remaining coefficients $a^{out}$ and $b^{in}$ are determined
by requiring a continuity of the pressure and flow field at $r=r_{0}$, leading to 
\begin{equation}
a^{out}=-\frac{1}{2}E\zeta_{0}r_{0}^{2}\ln\left|\frac{1}{2}E\zeta_{0}\right|,\quad b^{in}=-\frac{1}{2}E\zeta_{0}\left(\ln\left|\frac{1}{2}E\zeta_{0}\right|-\frac{1}{2}\right).\label{aout and bin coefficients}
\end{equation}

Utilizing (\ref{First order depth averaged velocity}), (\ref{RadialSymDistribution})
and (\ref{General solution FO NN p(r,th) dipole})-(\ref{aout and bin coefficients})
provides a closed-form expressions for the pressure and corresponding flow field at first order

\begin{equation}
p^{(1)}(r,\theta)=\left\{\begin{array}{ll}
\displaystyle{\frac{1}{4}\frac{E\zeta_{0}r_{0}^{2}}{r}\left(1-2\ln\left(\frac{1}{2}\frac{E\zeta_{0}r_{0}}{r}\right)\right)\cos(\theta)} & r>r_{0}\\
\displaystyle{\frac{1}{4}E\zeta_{0}\left(1-2\ln\left(\frac{1}{2}E\zeta_{0}\right)\right)r\cos(\theta)} & r<r_{0},
\end{array}\label{General solution FO p(r,th) dipole}\right.
\end{equation}
\begin{equation}
\left\langle \boldsymbol{u}_{\Vert}\right\rangle ^{(1)}=\left\{\begin{array}{ll}
\displaystyle{\frac{1}{4}\frac{E\zeta_{0}r_{0}^{2}}{r^{2}}\left[\left(-1+2\ln\left(\frac{r_{0}}{r}\right)\right)\cos(\theta)\boldsymbol{\hat{r}}+\left(1+2\ln\left(\frac{r_{0}}{r}\right)\right)\sin(\theta)\boldsymbol{\hat{\theta}}\right]} & r>r_{0}\vspace{1mm}\\
\displaystyle{\frac{1}{4}E\zeta_{0}\left[-\cos(\theta)\boldsymbol{\hat{r}}+\sin(\theta)\boldsymbol{\hat{\theta}}\right]} & r<r_{0}.
\end{array}\label{FO (only NN) velocity dipole}\right.
\end{equation}
Similarly to one dimensional case, the maximal pressure of shear-thinning
fluids, attained at $r=r_{0}$, can be either greater or less than
the pressure of shear-thickening fluids depending on the value of
$E\zeta_{0}$, where the transition occurs at $E\zeta_{0cr}\backsimeq3.29$. 

Figure 6 summarizes the effect of non-Newtonian behavior on the resulting
pressure and velocity magnitude in the Hele-Shaw cell for the problem
considered. Figures (6a) and (6b) present the pressure distribution
in the first quadrant for shear-thinning and shear-thickening fluids,
respectively. We note that, in order to provide clear comparison between
the cases, we chose to present the result for $n=0.5$ and $n=1.5$,
which for the previous 2D problem considered (see figure (5a)) bracketed
the range of $n$ values for which the error was less $10\,\%$. However,
since the current case has not analytical solution, the precise accuracy
of the solution at these $n$ values is unknown. The shear-thinning
fluid (figure (6a)) is characterized by greater spacing in the isobars,
compared to the shear-thickening fluid (figure (6b)). For further
clarification, figures (6c) and (6d) present the same isobar and equi-velocity
line, for three different values of the parameter $n$. This is also
shown quantitatively in figures (6e) showing the distance from the
origin over which the pressure decays to $10\,\%$ of its peak value,
as a function of $n$. The solid part of the line corresponds to the
range of $n$ lying between 0.5 and 1.75, where the asymptotic approximation
is assumed to be valid (according to the previous estimations) whereas
the dashed region may involve significant errors. Figure (6f) presents
a complementary view, indicating that the velocity magnitude decays
slower for larger values of $n$. 

\begin{figure}
\centerline{\includegraphics[scale=0.5]{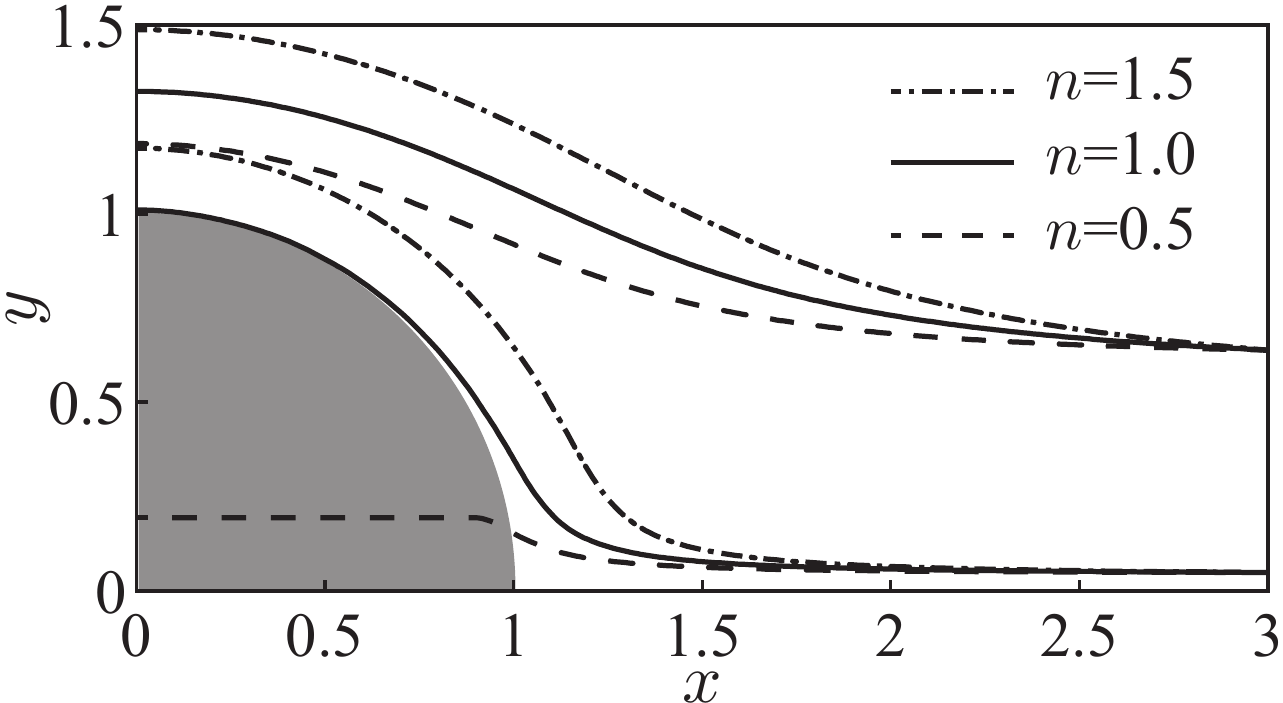}}
\caption{Effect of different non-Newtonian behavior on the resulting streamlines
(in first quadrant) due to a positive zeta potential, $\zeta_{0}/2$,
inside the spot of radius $r_{0}$ and negative zeta potential, $-\zeta_{0}/2$,
outside the spot. In the Newtonian case (solid lines) the streamlines
bend around the spot without entering it. The streamlines of the shear-thinning
fluid ($n=0.5$, dashed lines) are diverted toward the disk, while
the streamlines of the shear-thickening fluid ($n=1.5$, dashed-dot
lines) divert away from it. All calculations were performed using
$E\zeta_{0}=4$ and $r_{0}=1$.}
\end{figure}

To highlight the effect of different non-Newtonian behavior on the
resulting flow field, we take advantage of the uniform flow field
in the inner region and of the fact that by adding a constant value
to zeta potential the resulting pressure remains unaffected. Adding a bias value of $-\zeta_{0}/2$ to the
zeta potential everywhere in the domain, leads to zero net flow in
the spot for the Newtonian fluid and the resulting streamlines curve
around the spot without penetrating it. Figure 7 presents the resulting
streamlines for shear-thinning, Newtonian and shear-thickening fluids.
The incoming black dashed streamlines, which represent shear-thinning
behavior, are diverted towards the disk penetrating to it, due to
negative net flow inside the spot. On the other hand, the incoming
black dashed-dot streamlines, which represent shear-thinning behavior,
are diverted away from the disk, owing to positive net flow inside
the spot. This is consistent with (\ref{FO (only NN) velocity dipole})
which predicts a positive velocity field, $\left(1/4\right)\left|\delta\right|E\zeta_{0}\boldsymbol{\hat{x}}$,
for shear-thickening fluids and a negative velocity field, $-\left(1/4\right)\left|\delta\right|E\zeta_{0}\boldsymbol{\hat{x}}$,
for shear-thinning fluids inside the spot.

\section{Concluding remarks}

In this work, we studied the flow and pressure fields of non-Newtonian
fluids in a Hele-Shaw configuration, subjected to non-uniform EOF.
Using a power-law constitutive model, and under a depletion regime,
we derived a p-Poisson governing equation for the pressure, as well
as its asymptotic approximation for weakly non-Newtonian fluids. Our
analysis revealed that the asymptotic approximation may be applied
for values of $n$ between 0.5 and 1.75, while maintaining errors
below $10\,\%$.

We obtained that the maximal pressure due to shear-thickening fluids
is greater than the pressure of shear-thinning fluids only for sufficiently
large values of $E\zeta_{0}$. On the other hand, the velocity magnitude
of shear-thickening fluids is greater compared to shear-thinning fluids
for all values of $E\zeta_{0}$. The velocity field behavior depends
on the boundary conditions, where for a Dirichlet-type boundary condition
on the pressure (see $\mathsection$ 4.1 and $\mathsection$ 5.2),
the resulting flow field depends on parameter $n$, while for a Neumann-type
boundary condition it is independent of $n$, due to an implicitly
prescribed flux (see $\mathsection$ 5.1). We also showed that using
common non-Newtonian fluids with wall depletion properties (e.g. polymer
solutions) allows to increase the pressure resulting from non-uniform
surface patterning by more than two orders of magnitude compared to
Newtonian fluids. 

In our study we used a simple power-law constitutive model that is
valid only provided the shear rate $\dot{\gamma}$ (or $E\zeta_{0}$)
is within specific bounds. While a more general Carreau constitutive
model \citep{bird1987dynamics} can be applied to describe the entire range
of shear rates, its use greatly increases the complexity of the analytical
approach. In addition, we here neglected any viscoelastic effects.
The use of power-law models holds well for polymer solutions such
as CMC and PAA at low concentrations, for which viscoelastic effects
are negligible compared to viscous (shear-thinning) effects \citep{lin1995effects,ghannam1997rheological,roberts2001new,kim2003rheological}.
However, at higher polymer concentration viscoelastic effects are
apparent, and more complex constitutive model such as PTT would be
required. 

The ability to predict the velocity and pressure fields of non-Newtonian fluids subjected to non-uniform EOF, together with the inherent pressure increase obtained for depletion-regime fluids, opens the door to enhancement of existing electro-kinetic devices such as electroosmotic pumps, may lead to better control of dispersion processes, and to the development of new applications involving fluid-structure interaction.

\section*{Acknowledgments}
This project has received funding from the European Research Council (ERC) under the European Union's Horizon 2020 Research and Innovation Programme, grant agreement No. 678734 (MetamorphChip). We gratefully acknowledge supported by the Israel Science Foundation (grant No. 818/13).

\appendix
\section{Dimensional governing equations}

Here we summarize the main governing equations in a dimensional form.
The corresponding dimensional in-plane and perpendicular velocities are 
\begin{subequations}
\begin{equation}
\boldsymbol{\tilde{u}}_{||}=\frac{n}{n+1}\frac{\tilde{h}^{1+\frac{1}{n}}}{\tilde{\mu}_{eff}^{\frac{1}{n}}}\left|\boldsymbol{\tilde{\nabla}}_{\Vert}\tilde{p}\right|^{\frac{1}{n}-1}\boldsymbol{\tilde{\nabla}}_{\Vert}\tilde{p}\left(\left|\frac{\tilde{z}}{\tilde{h}}\right|^{1+\frac{1}{n}}-1\right)-\frac{\tilde{\varepsilon}\tilde{\zeta}(\tilde{x},\tilde{y})\tilde{\boldsymbol{E}}_{\Vert}}{\tilde{\eta}_{N}},\label{C1}
\end{equation}
\begin{equation}
\boldsymbol{\tilde{u}}_{\bot}=\frac{n}{n+1}\tilde{z}\left(1-\left|\frac{\tilde{z}}{\tilde{h}}\right|^{1+\frac{1}{n}}\right)\left[\boldsymbol{\tilde{\nabla}}_{\Vert}\left(-\frac{\tilde{\varepsilon}\tilde{\zeta}(\tilde{x},\tilde{y})}{\tilde{\eta}_{N}}\right)\cdot\tilde{\boldsymbol{E}}_{\Vert}\right]\boldsymbol{\hat{z}},\label{C2}
\end{equation}
\end{subequations}respectively. 
With the definition for depth-averaged
velocity, $\left\langle \boldsymbol{\tilde{u}}_{\Vert}\right\rangle =\left(1/\tilde{h}\right)\intop_{\tilde{z}=0}^{\tilde{z}=\tilde{h}}\boldsymbol{\tilde{u}}_{||}d\tilde{z},$
and using (\ref{C1}) yields 
\begin{equation}
\left\langle \boldsymbol{\tilde{u}}_{\Vert}\right\rangle =-\left(\frac{n}{2n+1}\right)\frac{\tilde{h}^{1+\frac{1}{n}}}{\tilde{\mu}_{eff}^{\frac{1}{n}}}\left|\boldsymbol{\tilde{\nabla}}_{\Vert}\tilde{p}\right|^{\frac{1}{n}-1}\boldsymbol{\tilde{\nabla}}_{\Vert}\tilde{p}-\frac{\tilde{\varepsilon}\tilde{\zeta}(\tilde{x},\tilde{y})\tilde{\boldsymbol{E}}_{\Vert}}{\tilde{\eta}_{N}},\label{C3}
\end{equation}
while applying two dimensional divergence to (\ref{C3}) results in dimensional governing p-Poisson equation for the pressure
\begin{equation}
\boldsymbol{\tilde{\nabla}}_{\Vert}\cdot\left(\left|\boldsymbol{\tilde{\nabla}}_{\Vert}\tilde{p}\right|^{\frac{1}{n}-1}\boldsymbol{\tilde{\nabla}}_{\Vert}\tilde{p}\right)=-\left(\frac{2n+1}{n}\right)\frac{\tilde{\mu}_{eff}^{\frac{1}{n}}}{\tilde{\eta}_{N}}\frac{\tilde{\varepsilon}}{\tilde{h}^{1+\frac{1}{n}}}\tilde{\boldsymbol{E}}_{\Vert}\cdot\boldsymbol{\tilde{\nabla}}_{\Vert}\tilde{\zeta}.\label{C4}
\end{equation}
We note that $\tilde{\mu}_{eff}$ here is measured in units of $\mathrm{Pa\,s^{n}}$
and depends on the value of $n$. For weakly non-Newtonian behavior
we define $n=1-\delta$, and the dimensional leading
and first orders equations for the pressure (\ref{Leading order pressure})
and (\ref{Leading order pressure}) read, respectively\begin{subequations}
\begin{equation}
O(1):\quad\tilde{\nabla}_{\Vert}^{2}\tilde{p}^{(0)}=\left(-\frac{\tilde{p}^{*}}{\tilde{E}^{*}\tilde{\zeta}^{*}\tilde{l}}\right)\tilde{\boldsymbol{E}}_{\Vert}\cdot\boldsymbol{\tilde{\nabla}}_{\Vert}\tilde{\zeta},\label{C5}
\end{equation}
\begin{equation}
O(\delta):\quad\tilde{\nabla}_{\Vert}^{2}\tilde{p}^{(1)}=-\boldsymbol{\tilde{\nabla}}_{\Vert}\cdot\left(\ln\left|\frac{\tilde{l}}{\tilde{p}^{*}}\boldsymbol{\tilde{\nabla}}_{\Vert}p^{(0)}\right|\boldsymbol{\tilde{\nabla}}_{\Vert}p^{(0)}\right),\label{C6}
\end{equation}
\end{subequations}
where $\tilde{p}^{*}$ is a characteristic pressure 
\begin{equation}
\tilde{p}^{*}=\left(\frac{2n+1}{n}\right)^{n}\frac{\tilde{\mu}_{eff}}{\epsilon^{n+1}\tilde{l}^{n}}\left(\frac{\tilde{\varepsilon}\left|\tilde{\zeta}^{*}\right|\tilde{E}^{*}}{\tilde{\eta}_{N}}\right)^{n}.\label{C7}
\end{equation}

\bibliographystyle{jfm}
\bibliography{nn}

\begin{thebibliography}{49}
\expandafter\ifx\csname natexlab\endcsname\relax\def\natexlab#1{#1}\fi
\def\au#1{#1} \def\ed#1{#1} \def\yr#1{#1}\def\at#1{#1}\def\jt#1{\textit{#1}}
  \def\bt#1{#1}\def\bvol#1{\textbf{#1}} \def\vol#1{#1} \def\pg#1{#1}
  \def\publ#1{#1}\def\arxiv#1{#1}\def\org#1{#1}\def\st#1{\textit{#1}}

\bibitem[Abramowitz \& Stegun(1964)]{abramowitz1964handbook}
{\sc \au{Abramowitz, M.} \& \au{Stegun, I.~A.}} \yr{1964} {\em Handbook of
  Mathematical Functions\/}.  \publ{Dover}.

\bibitem[Afonso {\em et~al.\/}(2009)Afonso, Alves \&
  Pinho]{afonso2009analytical}
{\sc \au{Afonso, A.~M.}, \au{Alves, M.~A.} \& \au{Pinho, F.~T.}} \yr{2009}
  \at{Analytical solution of mixed electro-osmotic/pressure driven flows of
  viscoelastic fluids in microchannels}.  \jt{J. Non-Newtonian Fluid Mech.}
  \bvol{159}~(1),  \pg{50--63}.

\bibitem[Afonso {\em et~al.\/}(2011)Afonso, Alves \& Pinho]{afonso2011electro}
{\sc \au{Afonso, A.~M.}, \au{Alves, M.~A.} \& \au{Pinho, F.~T.}} \yr{2011}
  \at{Electro-osmotic flow of viscoelastic fluids in microchannels under
  asymmetric zeta potentials}.  \jt{J. Eng. Math.}  \bvol{71}~(1),
  \pg{15--30}.

\bibitem[Afonso {\em et~al.\/}(2013)Afonso, Alves \&
  Pinho]{afonso2013analytical}
{\sc \au{Afonso, A.~M.}, \au{Alves, M.~A.} \& \au{Pinho, F.~T.}} \yr{2013}
  \at{Analytical solution of two-fluid electro-osmotic flows of viscoelastic
  fluids}.  \jt{J. Colloid Interface Sci.}  \bvol{395},  \pg{277--286}.

\bibitem[Ajdari(1995)]{ajdari1995electro}
{\sc \au{Ajdari, A.}} \yr{1995}  \at{Electro-osmosis on inhomogeneously charged
  surfaces}.  \jt{Phys. Rev. Lett.}  \bvol{75}~(4),  \pg{755}.

\bibitem[Ajdari(1996)]{ajdari1996generation}
{\sc \au{Ajdari, A.}} \yr{1996}  \at{Generation of transverse fluid currents
  and forces by an electric field: Electro-osmosis on charge-modulated and
  undulated surfaces}.  \jt{Phys. Rev. E}  \bvol{53}~(5),  \pg{4996}.

\bibitem[Ajdari(2001)]{ajdari2001transverse}
{\sc \au{Ajdari, A.}} \yr{2001}  \at{Transverse electrokinetic and microfluidic
  effects in micropatterned channels: Lubrication analysis for slab
  geometries}.  \jt{Phys. Rev. E}  \bvol{65}~(1),  \pg{016301}.

\bibitem[Anderson \& Idol(1985)]{anderson1985electroosmosis}
{\sc \au{Anderson, J.~L.} \& \au{Idol, W.~K.}} \yr{1985}  \at{Electroosmosis
  through pores with nonuniformly charged walls}.  \jt{Chem. Engng. Commun.}
  \bvol{38}~(3-6),  \pg{93--106}.

\bibitem[Aronsson \& Janfalk(1992)]{aronsson1992hele}
{\sc \au{Aronsson, G.} \& \au{Janfalk, U.}} \yr{1992}  \at{On hele-shaw flow of
  power-law fluids}.  \jt{Eur. J. Appl. Maths}  \bvol{3}~(04),  \pg{343--366}.

\bibitem[Babaie {\em et~al.\/}(2011)Babaie, Sadeghi \&
  Saidi]{babaie2011combined}
{\sc \au{Babaie, A.}, \au{Sadeghi, A.} \& \au{Saidi, M.~H.}} \yr{2011}
  \at{Combined electroosmotically and pressure driven flow of power-law fluids
  in a slit microchannel}.  \jt{J. Non-Newtonian Fluid Mech.}  \bvol{166}~(14),
   \pg{792--798}.

\bibitem[Barnes(1995)]{barnes1995review}
{\sc \au{Barnes, H.~A.}} \yr{1995}  \at{A review of the slip (wall depletion)
  of polymer solutions, emulsions and particle suspensions in viscometers: its
  cause, character, and cure}.  \jt{J. Non-Newtonian Fluid Mech.}
  \bvol{56}~(3),  \pg{221--251}.

\bibitem[Berli(2010)]{berli2010output}
{\sc \au{Berli, C. L.~A.}} \yr{2010}  \at{Output pressure and efficiency of
  electrokinetic pumping of non-{Newtonian} fluids}.  \jt{Microfluid Nanofluid}
   \bvol{8}~(2),  \pg{197--207}.

\bibitem[Berli \& Olivares(2008)]{berli2008electrokinetic}
{\sc \au{Berli, C. L.~A.} \& \au{Olivares, M.~L.}} \yr{2008}
  \at{Electrokinetic flow of non-{Newtonian} fluids in microchannels}.  \jt{J.
  Colloid Interface Sci.}  \bvol{320}~(2),  \pg{582--589}.

\bibitem[Bird {\em et~al.\/}(1987)Bird, Armstrong \&
  Hassager]{bird1987dynamics}
{\sc \au{Bird, R.~B.}, \au{Armstrong, R.~C.} \& \au{Hassager, O.}} \yr{1987}
  {\em Dynamics of polymeric liquids. Volume 1: Fluid Mechanics\/}.  \publ{2nd
  ed. John Wiley and Sons}.

\bibitem[Boyko {\em et~al.\/}(2015)Boyko, Rubin, Gat \&
  Bercovici]{boyko2015flow}
{\sc \au{Boyko, E.}, \au{Rubin, S.}, \au{Gat, A.~D.} \& \au{Bercovici, M.}}
  \yr{2015}  \at{Flow patterning in {Hele-Shaw} configurations using
  non-uniform electro-osmotic slip}.  \jt{Phys. Fluids}  \bvol{27}~(10),
  \pg{102001}.

\bibitem[Das \& Chakraborty(2006)]{das2006analytical}
{\sc \au{Das, S.} \& \au{Chakraborty, S.}} \yr{2006}  \at{Analytical solutions
  for velocity, temperature and concentration distribution in electroosmotic
  microchannel flows of a non-{Newtonian} bio-fluid}.  \jt{Anal. Chim. Acta}
  \bvol{559}~(1),  \pg{15--24}.

\bibitem[Derjaguin {\em et~al.\/}(1961)Derjaguin, Dukhin \&
  Korotkova]{derjaguin1961diffusiophoresis}
{\sc \au{Derjaguin, B.~V.}, \au{Dukhin, S.~S.} \& \au{Korotkova, A.~A.}}
  \yr{1961}  \at{Diffusiophoresis in electrolyte solutions and its role in the
  mechanism of film formation from rubber latexes by the method of ionic
  deposition}.  \jt{Kolloidn. Zh.}  \bvol{23}~(1),  \pg{53}.

\bibitem[Derjaguin {\em et~al.\/}(1993)Derjaguin, Dukhin \&
  Korotkova]{derjaguin1993diffusiophoresis}
{\sc \au{Derjaguin, B.~V.}, \au{Dukhin, S.~S.} \& \au{Korotkova, A.~A.}}
  \yr{1993}  \at{Diffusiophoresis in electrolyte solutions and its role in the
  mechanism of the formation of films from caoutchouc latexes by the ionic
  deposition method}.  \jt{Prog. Surf. Sci.}  \bvol{43}~(1),  \pg{153--158}.

\bibitem[Dhinakaran {\em et~al.\/}(2010)Dhinakaran, Afonso, Alves \&
  Pinho]{dhinakaran2010steady}
{\sc \au{Dhinakaran, S.}, \au{Afonso, A.~M.}, \au{Alves, M.~A.} \& \au{Pinho,
  F.~T.}} \yr{2010}  \at{Steady viscoelastic fluid flow between parallel plates
  under electro-osmotic forces: {Phan-Thien-Tanner} model}.  \jt{J. Colloid
  Interface Sci.}  \bvol{344}~(2),  \pg{513--520}.

\bibitem[Erickson \& Li(2002)]{erickson2002influence}
{\sc \au{Erickson, D.} \& \au{Li, D.}} \yr{2002}  \at{Influence of surface
  heterogeneity on electrokinetically driven microfluidic mixing}.
  \jt{Langmuir}  \bvol{18}~(5),  \pg{1883--1892}.

\bibitem[Erickson \& Li(2003)]{erickson2003three}
{\sc \au{Erickson, D.} \& \au{Li, D.}} \yr{2003}  \at{Three-dimensional
  structure of electroosmotic flow over heterogeneous surfaces}.  \jt{J. Phys.
  Chem.}  \bvol{107}~(44),  \pg{12212--12220}.

\bibitem[Ghannam \& Esmail(1997)]{ghannam1997rheological}
{\sc \au{Ghannam, M.~T.} \& \au{Esmail, M.~N.}} \yr{1997}  \at{Rheological
  properties of carboxymethyl cellulose}.  \jt{J. Appl. Polymer Sci.}
  \bvol{64}~(2),  \pg{289--301}.

\bibitem[Ghosh \& Chakraborty(2015)]{ghosh2015electroosmosis}
{\sc \au{Ghosh, U.} \& \au{Chakraborty, S.}} \yr{2015}  \at{Electroosmosis of
  viscoelastic fluids over charge modulated surfaces in narrow confinements}.
  \jt{Phys. Fluids}  \bvol{27}~(6),  \pg{062004}.

\bibitem[Hunter(2000)]{hunter2001foundations}
{\sc \au{Hunter, R.~J.}} \yr{2000} {\em Foundations of colloid science\/}.
  \publ{Oxford University Press}.

\bibitem[Irgens(2014)]{irgens2014rheology}
{\sc \au{Irgens, F.}} \yr{2014} {\em Rheology and Non-Newtonian Fluids\/}.
  \publ{Springer}.

\bibitem[Khair \& Squires(2008{\natexlab{{\em a\/}}})]{khair2008fundamental}
{\sc \au{Khair, A.~S.} \& \au{Squires, T.~M.}} \yr{2008{\natexlab{{\em a\/}}}}
  \at{Fundamental aspects of concentration polarization arising from nonuniform
  electrokinetic transport}.  \jt{Phys. Fluids}  \bvol{20}~(8),  \pg{087102}.

\bibitem[Khair \& Squires(2008{\natexlab{{\em b\/}}})]{khair2008surprising}
{\sc \au{Khair, A.~S.} \& \au{Squires, T.~M.}} \yr{2008{\natexlab{{\em b\/}}}}
  \at{Surprising consequences of ion conservation in electro-osmosis over a
  surface charge discontinuity}.  \jt{J. Fluid Mech.}  \bvol{615},
  \pg{323--334}.

\bibitem[Kim {\em et~al.\/}(2003)Kim, Song, Lee \& Park]{kim2003rheological}
{\sc \au{Kim, J.-Y.}, \au{Song, J.-Y.}, \au{Lee, E.-J.} \& \au{Park, S.-K.}}
  \yr{2003}  \at{Rheological properties and microstructures of carbopol gel
  network system}.  \jt{Colloid Polym. Sci.}  \bvol{281}~(7),  \pg{614--623}.

\bibitem[Lin \& Ko(1995)]{lin1995effects}
{\sc \au{Lin, C.-X.} \& \au{Ko, S.-Y.}} \yr{1995}  \at{Effects of temperature
  and concentration on the steady shear properties of aqueous solutions of
  carbopol and cmc}.  \jt{Int. Commun. Heat Mass Transfer}  \bvol{22}~(2),
  \pg{157--166}.

\bibitem[Lyklema(1995)]{lyklemafundamentals}
{\sc \au{Lyklema, J.}} \yr{1995} {\em Fundamentals of Interface and Colloid
  Science. Volume II: Solid-Liquid Interfaces\/}.  \publ{Academic}.

\bibitem[Ng \& Qi(2014)]{ng2014electroosmotic}
{\sc \au{Ng, C.} \& \au{Qi, C.}} \yr{2014}  \at{Electroosmotic flow of a
  power-law fluid in a non-uniform microchannel}.  \jt{J. of Non-Newtonian
  Fluid Mech.}  \bvol{208},  \pg{118--125}.

\bibitem[Olivares {\em et~al.\/}(2009)Olivares, Vera-Candioti \&
  Berli]{olivares2009eof}
{\sc \au{Olivares, M.~L.}, \au{Vera-Candioti, L.} \& \au{Berli, C. L.~A.}}
  \yr{2009}  \at{The {EOF} of polymer solutions}.  \jt{Electrophoresis}
  \bvol{30}~(5),  \pg{921--928}.

\bibitem[Paul(2008)]{paul2008electrokinetic}
{\sc \au{Paul, P.~H.}} \yr{2008} Electrokinetic device employing a
  non-{Newtonian} liquid. US Patent, US7429317.

\bibitem[Prieve {\em et~al.\/}(1984)Prieve, Anderson, Ebel \&
  Lowell]{prieve1984motion}
{\sc \au{Prieve, D.~C.}, \au{Anderson, J.~L.}, \au{Ebel, J.~P.} \& \au{Lowell,
  M.~E.}} \yr{1984}  \at{Motion of a particle generated by chemical gradients.
  {Part 2.} {Electrolytes}}.  \jt{J. Fluid Mech.}  \bvol{148},  \pg{247--269}.

\bibitem[Qi \& Ng(2015)]{qi2015electroosmotic1}
{\sc \au{Qi, C.} \& \au{Ng, C.}} \yr{2015}  \at{Electroosmotic flow of a
  power-law fluid in a slit microchannel with gradually varying channel height
  and wall potential}.  \jt{Eur. J. of Mech.-B/Fluids}  \bvol{52},
  \pg{160--168}.

\bibitem[Qian \& Bau(2002)]{qian2002chaotic}
{\sc \au{Qian, S.} \& \au{Bau, H.~H.}} \yr{2002}  \at{A chaotic electroosmotic
  stirrer}.  \jt{Anal.Chem.}  \bvol{74}~(15),  \pg{3616--3625}.

\bibitem[Roberts \& Barnes(2001)]{roberts2001new}
{\sc \au{Roberts, G.~P.} \& \au{Barnes, H.~A.}} \yr{2001}  \at{New measurements
  of the flow-curves for carbopol dispersions without slip artefacts}.
  \jt{Rheol. Acta}  \bvol{40}~(5),  \pg{499--503}.

\bibitem[Rubin {\em et~al.\/}(2016)Rubin, Tulchinky, Gat \&
  Bercovici]{rubin2016}
{\sc \au{Rubin, S.}, \au{Tulchinky, A.}, \au{Gat, A.} \& \au{Bercovici, M.}}
  \yr{2016}  \at{Elastic deformations driven by non-uniform lubrication flows}.
   \jt{under consideration in J. Fluid Mech.} .

\bibitem[Sousa {\em et~al.\/}(2011)Sousa, Afonso, Pinho \&
  Alves]{sousa2011effect}
{\sc \au{Sousa, J.~J.}, \au{Afonso, A.~M.}, \au{Pinho, F.~T.} \& \au{Alves,
  M.~A.}} \yr{2011}  \at{Effect of the skimming layer on
  electroosmotic-{Poiseuille} flows of viscoelastic fluids}.  \jt{Microfluid
  Nanofluid}  \bvol{10}~(1),  \pg{107--122}.

\bibitem[Stroock {\em et~al.\/}(2000)Stroock, Weck, Chiu, Huck, Kenis,
  Ismagilov \& Whitesides]{stroock2000patterning}
{\sc \au{Stroock, A.~D.}, \au{Weck, M.}, \au{Chiu, D.~T.}, \au{Huck, W. T.~S.},
  \au{Kenis, P. J.~A.}, \au{Ismagilov, R.~F.} \& \au{Whitesides, G.~M.}}
  \yr{2000}  \at{Patterning electro-osmotic flow with patterned surface
  charge}.  \jt{Phys. Rev. Lett.}  \bvol{84}~(15),  \pg{3314}.

\bibitem[Tang {\em et~al.\/}(2009)Tang, Li, He \& Tao]{tang2009electroosmotic}
{\sc \au{Tang, G.~H.}, \au{Li, X.~F.}, \au{He, Y.~L.} \& \au{Tao, W.~Q.}}
  \yr{2009}  \at{Electroosmotic flow of non-{Newtonian} fluid in
  microchannels}.  \jt{J. Non-Newtonian Fluid Mech.}  \bvol{157}~(1),
  \pg{133--137}.

\bibitem[Vakili {\em et~al.\/}(2012)Vakili, Sadeghi, S. \&
  Mozafari]{vakili2012electrokinetically}
{\sc \au{Vakili, M.~A.}, \au{Sadeghi, A.}, \au{S., Mohammad~H.} \&
  \au{Mozafari, A.~A.}} \yr{2012}  \at{Electrokinetically driven fluidic
  transport of power-law fluids in rectangular microchannels}.  \jt{Colloids
  Surf., A}  \bvol{414},  \pg{440--456}.

\bibitem[Vasu \& De(2010)]{vasu2010electroosmotic}
{\sc \au{Vasu, N.} \& \au{De, S.}} \yr{2010}  \at{Electroosmotic flow of
  power-law fluids at high zeta potentials}.  \jt{Colloids Surf., A}
  \bvol{368}~(1),  \pg{44--52}.

\bibitem[Zhang {\em et~al.\/}(2006)Zhang, He \& Liu]{zhang2006electro}
{\sc \au{Zhang, J.}, \au{He, G.} \& \au{Liu, F.}} \yr{2006}  \at{Electroosmotic
  flow and mixing in heterogeneous microchannels}.  \jt{Phys. Rev. E}
  \bvol{73}~(5),  \pg{056305}.

\bibitem[Zhao \& Yang(2010)]{zhao2010nonlinear}
{\sc \au{Zhao, C.} \& \au{Yang, C.}} \yr{2010}  \at{Nonlinear {Smoluchowski}
  velocity for electroosmosis of power-law fluids over a surface with arbitrary
  zeta potentials}.  \jt{Electrophoresis}  \bvol{31}~(5),  \pg{973--979}.

\bibitem[Zhao \& Yang(2011)]{zhao2011exact}
{\sc \au{Zhao, C.} \& \au{Yang, C.}} \yr{2011}  \at{An exact solution for
  electroosmosis of non-{Newtonian} fluids in microchannels}.  \jt{J.
  Non-Newtonian Fluid Mech.}  \bvol{166}~(17),  \pg{1076--1079}.

\bibitem[Zhao \& Yang(2013{\natexlab{{\em a\/}}})]{zhao2013electrokinetics}
{\sc \au{Zhao, C.} \& \au{Yang, C.}} \yr{2013{\natexlab{{\em a\/}}}}
  \at{Electrokinetics of non-{Newtonian} fluids: a review}.  \jt{Adv. Colloid
  and Interface Sci.}  \bvol{201},  \pg{94--108}.

\bibitem[Zhao \& Yang(2013{\natexlab{{\em b\/}}})]{zhao2013electroosmotic}
{\sc \au{Zhao, C.} \& \au{Yang, C.}} \yr{2013{\natexlab{{\em b\/}}}}
  \at{Electroosmotic flows of non-{Newtonian} power-law fluids in a cylindrical
  microchannel}.  \jt{Electrophoresis}  \bvol{34}~(5),  \pg{662--667}.

\bibitem[Zhao {\em et~al.\/}(2008)Zhao, Zholkovskij, Masliyah \&
  Yang]{zhao2008analysis}
{\sc \au{Zhao, C.}, \au{Zholkovskij, E.}, \au{Masliyah, J.~H.} \& \au{Yang,
  C.}} \yr{2008}  \at{Analysis of electroosmotic flow of power-law fluids in a
  slit microchannel}.  \jt{J. Colloid Interface Sci.}  \bvol{326}~(2),
  \pg{503--510}.

\end{thebibliography}
\end{document}